\documentclass[aps,pra,preprint,tightenlines,groupedaddress]{revtex4-1}

\usepackage{graphicx} \usepackage[T1]{fontenc}
\usepackage[latin1]{inputenc} \usepackage[english]{babel}
\usepackage{amsmath,amsfonts,amssymb}
 \usepackage{array}
\renewcommand{\d}{\textrm{d}} \newcommand{\F}{\textrm{F}}

\newcommand{\bbraket}[2]{\bigl\langle{#1}\bigr|{#2}\bigr\rangle}

\newcommand{\bracket}[3]{\left\langle{#1}\right.\!\left|{#2}\right|\!\left.{#3}\right\rangle}
\newcommand{\bbracket}[3]{\bigl\langle{#1}\bigr|{#2}\bigl|{#3}\bigr\rangle}

\newcommand{\bket}[1]{\bigl|{#1}\bigl\rangle}
\begin{document}
% Use the \preprint command to place your local
% institutional report number in the upper righthand corner
% of the title page in preprint mode.  Multiple \preprint
% commands are allowed.  Use the 'preprintnumbers' class
% option to override journal defaults to display numbers if
% necessary \preprint{}

% Title of paper
\title{Convergence Properties of the Effective Theory for
  Trapped Bosons} \author{S.\ T\"olle}
\email{toelle@hiskp.uni-bonn.de} \author{H.-W.\ Hammer}
\email{hammer@hiskp.uni-bonn.de} \author{B.\ Ch.\ Metsch}
\email{metsch@hiskp.uni-bonn.de}

% \homepage[]{Your web page} \thanks{} \altaffiliation{}
\affiliation{Helmholtz-Institut f\"ur Strahlen- und
  Kernphysik and Bethe Center for Theoretical Physics,
  Universit\"at Bonn, 53115 Bonn, Germany}

% Collaboration name if desired (requires use of
% superscriptaddress option in
% \documentclass). \noaffiliation is required (may also be
% used with the \author command).  \collaboration can be
% followed by \email, \homepage, \thanks as well.
% \collaboration{} \noaffiliation

\date{\today}

\begin{abstract}
  We investigate few-boson systems with resonant
  interactions in a narrow harmonic trap within an effective
  theory framework. The size of the model space is
  identified with the effective theory cutoff.  In the
  universal regime, the interactions of the bosons can be
  approximated by contact interactions. We investigate the
  convergence properties of genuine and smeared contact
  interactions as the size of the model space is increased
  and present a detailed error analysis. The spectra for
  few-boson systems with up to 6 identical particles are
  calculated by combining extrapolations in the cutoff and
  in the smearing parameter.
\end{abstract}


% insert suggested keywords - APS authors don't need to do
% this \keywords{}

% \maketitle must follow title, authors, abstract, \pacs,
% and \keywords
\maketitle

% References should be done using the \cite, \ref,
% and \label commands
% \section{Approach \label{approach}}
\section{Introduction}

Few-body systems with resonant interactions show universal
properties independent of the details of the interaction at
short distances \cite{Braaten:2004rn}.  If the scattering
length $a$ is much larger than the range of the interaction
$r_0$, the underlying interaction can be approximated by an
effective theory with contact interactions.  In the two-body
system, the universal properties include the existence of a
shallow dimer with binding energy $E_2 \approx
-{\hbar^2}/{(m a^2)}\,$, if $a$ is positive. In the
three-body system, the Efimov effect then generates a
geometric spectrum of three-body bound states
\cite{Efimov-70}:
% ----------------------
\begin{eqnarray}
  E^{(n)}_3 = -(e^{-2\pi/s_0})^{n} \hbar^2 \kappa^2_* /m,\quad n=0,1,2,\ldots
  \label{kappa-star}
\end{eqnarray}
% ----------------------
where $s_0=1.00624\ldots$ and $\kappa_*$ is the binding
momentum of the lowest Efimov state. This spectrum shows a
discrete scaling symmetry which corresponds to a
renormalization group limit cycle \cite{Bedaque:1998kg}.
Recent theoretical studies have shown that these universal
properties also hold for the four-body system.  In
particular, there is a pair of universal four-body states
associated with every Efimov state
\cite{Hammer:2006ct,Stecher:2008}.  Moreover, universal
cluster states have been predicted in five-, six- and
possibly higher-body systems as well
\cite{Hanna:2006,Stecher:2011,Gattobigio:2011,Nicholson:2012}.
Using Feshbach resonances, the two-, three-, four- and
five-body states have been confirmed in a ultracold atomic
gases in a variety of experiments with different atom
species \cite{Ferlaino:2010, Chin:2010aa, Zenesini:2012}.

These experiments were carried out in a regime where the
influence of the trap on the few-body spectra could be
neglected.  However, the trap also offers new possibilities
to modify the properties of few-body systems. In particular,
a narrow harmonic confinement modifies the
spectrum. Microtraps with a small number of atoms offer the
opportunity to study the transition from few- to many-body
systems in a controlled environment.

Many studies have focused on fermions in the unitary regime
in a harmonic trap where the Efimov effect does not occur
(see Ref.~\cite{Blume:2012} for a review of this work).
Here, we focus on the case of spinless bosons.  The 2-body
problem of spinless bosons with contact interactions in an
isotropic harmonic trap was solved analytically by Busch
et~al.~\cite{Busch:1998}.  The corresponding 3-body problem
was first solved by Jonsell et~al.~\cite{Jonsell:2002}.
Werner and Castin calculated the complete 3-body spectrum in
the unitary limit of infinite scattering length and provided
a semi-analytic solution \cite{Werner:2006}.  More recent
work has focused on the universality of the two-body
spectrum \cite{Shea:2009,Zinner:2011}, the extension to
heteronuclear systems \cite{Chen:2011} and anisotropic traps
\cite{Idziaszek:2005,Liang:2008}, and the interpretation of
the scattering-length dependence of the level structure in
terms of the Zel$'$dovich effect \cite{Farrell:2012}.

In this work, we investigate universal few-body physics in
harmonic confinement for up to six bosons using an effective
theory framework.  Our strategy follows Stetcu
et~al.~\cite{Stetcu:2006ey,Stetcu:2009ic, Rotureau:2011vf},
where an effective theory for short-range nuclear forces in
the framework of the no-core shell model was formulated.
The effective interactions were defined within a finite
model space with a cutoff $N$ on the basis functions.  In
previous work \cite{toelle2010}, we have performed a
systematic study of the spectra of three- and four-body
systems in the vicinity of the unitary limit at leading
order in the effective theory.  In this paper, we extend our
calculations to six bodies.  The running of the coupling
constants for two- and three-body interactions with the
cutoff $N$ is investigated in detail.  Moreover, we
introduce smeared contact interactions.  We compare the
convergence of the effective theory as the size of the model
space is increased for exact and smeared contact
interactions and discuss different strategies to remove the
regulator.  Finally, spectra for $A$-boson systems with up
to six bodies are calculated using a combined extrapolation
in $N$ and the smearing parameter $\epsilon$.

\section{Theoretical Framework}

We investigate few-boson systems with resonant two-body
interactions characterized by a large S-wave scattering
length $a$ in a harmonic trap using an effective theory with
contact interactions.  The contact interactions lead to
ultraviolet divergences which have to be renormalized by
fixing one two-body energy (or, equivalently the scattering
length $a$) and a three-body energy
\cite{Bedaque:1998kg,Platter:2004qn}.  Since the trap only
modifies the infrared properties of the system, the
renormalization of the contact interactions is the same as
in free space. We stay at leading order in the effective
theory.  (See Refs.~\cite{Stetcu:2010,Rotureau:2010uz} for
an explicit calculation of effective range corrections for
two-component fermions.)  Higher-order corrections are
suppressed by $\kappa R$ where $\kappa=\sqrt{m|E|}$ is a
typical momentum and $R$ is the range of the
interaction. For broad Feshbach resonances $R$ is given by
the van der Waals length $l_{\textrm{vdW}}$. For trapped
systems, there are also corrections of order $R/\beta$,
where $\beta=\sqrt{\hbar/(m\omega)}$ is the oscillator
length.  As we will discuss below, there are additional
corrections if the two- and three-body renormalization
energies $E_2$ and $E_3$ are very different.

We implement this effective theory in a Hamiltonian
framework.  For a system of $A$ bosons in an isotropic
harmonic oscillator potential (HOP) at leading order the
Hamiltonian can be written as
\begin{align}
  H=\sum_{i=1}^{A}\left(\frac{\bigl|\vec{p}_i\bigr|^2}{2m}
    +\frac{1}{2}m\omega^2\bigl|\vec{x}_i\bigr|^2\right)
  +g^{(2)}\sum_{i<j}^A V_{i j}%\,\delta^{(3)}(\vec{x}_{ij})
  +g^{(3)}\sum_{i<j<k}^A W_{i j
    k}%\,\delta^{(3)}(\vec{x}_{ij})\delta^{(3)}(\vec{x}_{ik})
  \:,
  \label{eq:effham}
\end{align}
where %$\vec{x}_{ij}=\vec{x}_{i}-\vec{x}_j$,
$V$ and $W$ are two- and three-body contact interactions,
$m$ is the mass of the bosons, and $\omega$ is the trapping
frequency.  In the following, all lengths are rescaled with
the oscillator length $\beta=\sqrt{\hbar/(m\omega)}$ for
convenience. Since the interactions are Galilei invariant
and the HOP allows to separate the center of mass motion,
the problem is transformed to an $(A-1)$-body problem in
Jacobi coordinates $\vec{s}_n$ and a one body problem in
$\vec{R}$, where
\begin{align}
  \vec{s}_n&=\frac{1}{\beta\sqrt{(n+1)n}}
  (\sum_{j=1}^n\vec{x}_j-n\vec{x}_{n+1})\;,\qquad
  \vec{R}=\frac{1}{\beta\sqrt{A}}
  (\sum_{j=1}^A\vec{x}_{j})\;,
\end{align}
with the property
% \begin{align}
$ \sum_{j=1}^{A}|\vec{x}_j|^2= \beta^2
\left(\sum_{n=1}^{A-1}|\vec{s}_n|^2+|\vec{R}|^2\right)$.
% \end{align}

The Hamiltonian (\ref{eq:effham}) has to be regularized
since it is not self-adjoint. Hence the Hilbert space is
restricted to the model space consisting of the linear hull
of oscillator states $\{\bigotimes_{i=1}^{A-1}\bket{n_i l_i
  m_i}\}$ with eigenvalues
$\sum_{i=1}^{A-1}\hbar\omega(2n_i+l_i+\frac{3}{2})
\leq\hbar\omega(N+(A-1)\frac{3}{2})$
\cite{Stetcu:2007ms}. In order to match the results in the
model space to the full Hilbert space results, the coupling
constants $g^{(2)}$ and $g^{(3)}$ are renormalized (see
Sec.~\ref{sec:renormalization}). We identify $N$ with the
ultraviolet cutoff of our effective theory. The coupling
constants $g^{(2)}$ and $g^{(3)}$ are fixed by matching to a
two- and a three-body energy, respectively, and run with the
cutoff $N$.  In free space one usually uses a momentum
cutoff $\Lambda$ to regularize the resulting integral
equation and finds a linear dependence of observables on
$1/\Lambda$ in first order of perturbation theory. Because
of the relation $p\propto\sqrt{E}$ between momentum $p$ and
energy $E$, we expect errors in the eigenenergies in the
trap of order $1/\sqrt{N+(A-1)3/2}$.

In contrast to the momentum cutoff $\Lambda$ used in free
space, the regulator $N$ also implies an infrared cutoff. In
the literature, there are two definitions of the IR-cutoff
for an oscillator basis. Both the expression
$\beta/\sqrt{N+(A-1)3/2}$ \cite{Jurgenson:2011, Hagen:2010},
which quantifies the maximum size of structures that can be
captured in the given basis, or the oscillator length
$\beta$ itself \cite{Coon:2012} have been interpreted as an
IR-cutoff. In our calculations, the harmonic oscillator
represents a physical trap and thus $\beta$ is a physical
parameter.  The trap acts like a finite box that confines
the system. In this case, only the first definition is
sensible and we expect errors of order $1/\sqrt{N+(A-1)3/2}$
from the infrared cutoff. These errors have the same scaling
behavior with $N$ as the errors from the ultraviolet cutoff
and vanish for $N\rightarrow\infty$.

In our previous work, we found that the convergence to the
results in the full Hilbert space is slow due to the
irregularity of the $\delta$-distribution in the
parametrization of the genuine contact interactions
\cite{toelle2010}.  For the cutoff dependence to become
insignificant, values for $1/\sqrt{N}$ of order 0.1 or
smaller have to be reached.  For four- and higher-body
states such low values can in general not be obtained. The
eigenenergies are estimated by extrapolation of results for
finite $N$ to $N\rightarrow\infty$.

Here, we investigate a second possibility. The convergence
can be improved by smearing the contact interactions. We
keep the separability of the interaction and approximate the
contact interactions by narrow Gaussians with width
$\epsilon$.  More specifically, the separable two-body
contact interaction
\begin{align}
  \bbracket{\vec{s}}{V }{\vec{s\,}'}&
  =\delta^{(3)}(\vec{s})\delta^{(3)}(\vec{s}-\vec{s\,}')
  =\delta^{(3)}(\vec{s})\delta^{(3)}(\vec{s\,}')
\end{align}
is substituted by
\begin{align}
  \bbracket{\vec{s}}{V_\epsilon}{\vec{s\,}'}=
  (2\pi\epsilon^2)^{-3/2} e^{-|\vec{s}|^2/(2\epsilon^2)}\;
  (2\pi\epsilon^2)^{-3/2}
  e^{-|\vec{s\,}'|^2/(2\epsilon^2)}\,,
\end{align}
and the three-body contact interaction
\begin{align}
  \bbracket{\vec{s_1}, \vec{s_2}}{W }{\vec{s\,}_1',
    \vec{s\,}_2'}&
  % =\delta^{(3)}(\vec{s})\delta^{(3)}(\vec{s}-\vec{s\,}')
  =\delta^{(3)}(\vec{s_1})\delta^{(3)}(\vec{s\,}_1')\delta^{(3)}(\vec{s_2})
  \delta^{(3)}(\vec{s\,}_2')
\end{align}
is replaced by
\begin{align}
  \bbracket{\vec{s_1}, \vec{s_2}}{W_\epsilon }{\vec{s\,}_1',
    \vec{s\,}_2'}& =(2\pi\epsilon^2)^{-6}\;e^{-
    (|\vec{s}_1|^2+|\vec{s}_2|^2+
    |\vec{s\,}_1'|^2+|\vec{s\,}_2'|^2)/(2\epsilon^2)}\,.
\end{align}
The case of contact interactions is then recovered in the
limit $\epsilon\to0$.

\subsection{Matrix elements}\label{sec:matrixelements}

The eigenenergies are determined by diagonalization of the
Hamiltonian matrix in an angular-momentum-coupled basis of
oscillator states. For contact interactions as well as
smeared contact interactions one can easily find the matrix
elements for the oscillator basis functions
\begin{align}
  \phi_{n_1l_1m_1}(\vec{s}_1)= R_{n_1l_1}(|\vec{s}_1|)
  Y_{l_1m_1}(\Omega)= \bbraket{\vec{s}_{1}}{n_1l_1m_1}\;.
\end{align}

For two-body contact interactions, we have
\begin{align}
  \bbracket{n_1l_1m_1}{V } {n_1'l_1'm_1'}&=
  \phi_{n_10}(0)\phi_{n_1'0}(0)\,\delta_{l_10}\,\delta_{l_1'0}
  % =\frac{1}{\pi^{\frac{3}{2}}}
  % \sqrt{\frac{(2n_1+1)!!}{n_1!2^{n_1}}}
  % \sqrt{\frac{(2n_1'+1)!!}{n_1'!2^{n_1'}}}
  % \delta_{l_10}\delta_{l_1'0}
  \;,
\end{align}
where
\begin{align}
  \label{eq:phi0}
  \phi_{n0}(0)=
  \frac{1}{\pi^{3/4}}\left(\frac{\Gamma(n+\frac{3}{2})}
    {\Gamma(\frac{3}{2})\Gamma(n+1)}\right)^{1/2}=
  \left(\frac{(2n+1)!!}
    {\pi^{3/2}\,n!\,2^{n}}\right)^{1/2}\;.
\end{align}

For the smeared two-body contact interaction, we obtain
according to Appendix~\ref{sec:smear-cont-inter}
\begin{align}
  \bbracket{n_1l_1m_1}{V_\epsilon}{n_1'l_1'm_1'}&=
  \frac{1}{(1+\epsilon^2)^3}
  \Bigl(\frac{1-\epsilon^2}{1+\epsilon^2}\Bigr)^{n_1+n'_1}
  \phi_{n_10}(0) \phi_{n_1'0}(0)
  \delta_{l_10}\delta_{l_1'0}\;.
\end{align}

The three-body interactions are determined in a coupled
scheme, i.e.\ both angular momenta are coupled to a total
angular momentum with Clebsch-Gordan coefficients,
\begin{align}
  \bigl[\phi_{n_1l_1}(\vec{s}_1)\otimes
  \phi_{n_2l_2}(\vec{s}_2)\bigr]^L_M=
  \bbraket{\vec{s}_1,\vec{s}_2}{n_1l_1,n_2l_2,L M}\;.
\end{align}
\begin{figure}[ht]
  \begin{center}
    \parbox[c][5em][c]{0.4\linewidth}
    {\includegraphics[width=0.9\linewidth]{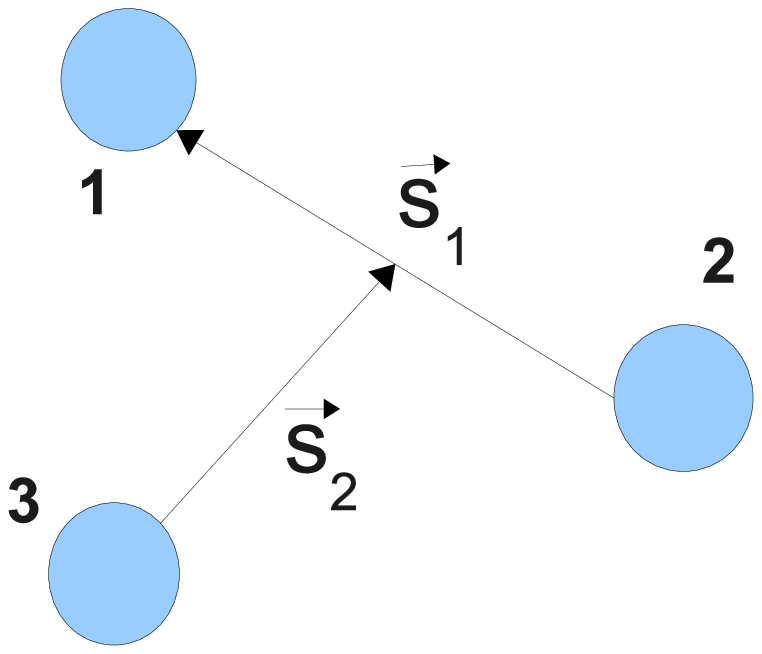}}
    \parbox[c][5em][c]{0.4\linewidth}
    {\begin{align*} \vec{s}_1&=\frac{1}{\beta\sqrt{2}}
        (\vec{x}_1-\vec{x}_{2})\;,\\
        \vec{s}_2&=\frac{1}{\beta\sqrt{6}}
        (\vec{x}_1+\vec{x}_{2}-2\vec{x}_3)\;,\\
        \vec{R}&=\frac{1}{\beta\sqrt{3}}
        (\vec{x}_1+\vec{x}_{2}+\vec{x}_3)\;.
      \end{align*}}
  \end{center}
  \vspace*{0.2cm}
  \caption{Set of Jacobi coordinates for the three-body
    system.}
  \label{fig:Jacobi3}
\end{figure}

We find for the three-body contact interactions in Jacobi
coordinates $\vec{s}_1$ and $\vec{s}_2$
\begin{align}
  \bbracket{n_1l_1,n_2l_2,L M} {W} {n_1'l_1',n_2'l_2',L'
    M'}&=
  \phi_{n_10}(0)\phi_{n_1'0}(0)\phi_{n_20}(0)\phi_{n_2'0}(0)
  \delta_{l_10}\delta_{l_1'0} \delta_{l_20}\delta_{l_2'0}\;,
\end{align}
and finally for the smeared three-body contact interaction
\begin{multline}
  \bbracket{n_1l_1,n_2l_2,L M} {W_\epsilon}
  {n_1'l_1',n_2'l_2',L' M'}=\\
  \frac{1}{(1+\epsilon^2)^6}
  \Bigl(\frac{1-\epsilon^2}{1+\epsilon^2}\Bigr)^{n_1+n_2+n_1'+n'_2}
  \phi_{n_10}(0) \phi_{n_20}(0) \phi_{n_1'0}(0)
  \phi_{n_2'0}(0) \delta_{l_10}\delta_{l_1'0}
  \delta_{l_2'0}\delta_{l_20}\;.
\end{multline}

\subsection{Renormalization}\label{sec:renormalization}

Next we discuss the renormalization of the Hamiltonian,
Eq.~(\ref{eq:effham}). For convenience, all energies are
given in units of $\hbar \omega$. The coupling constants
$g^{(2)}$ and $g^{(3)}$ depend on the regulator $N$. At each
value of $N$, i.e.\ in each model space, their values have
to be fixed by one two-body and one three-body
observable. All other observables are then predictions.

As the two-body input parameter, we use an eigenenergy
$E^{(2)}$ in the two-body spectrum with $L=0$. Using the
analytical solution derived by Busch and collaborators
\cite{Busch:1998},
\begin{align}
  \frac{\beta}{a}= \sqrt{2}\,\frac{\Gamma(3/4-E^{(2)}/2)}
  {\Gamma(1/4-E^{(2)}/2)}\;,
\end{align}
this energy $E^{(2)}$ is directly related to the S-wave
scattering length $a$.

The relation between $g^{(2)}$ and $E^{(2)}$ for a given $N$
can be found analytically by exploiting of the separability
of the interaction. We define
\begin{equation}
  f_{n\epsilon}\equiv
  \begin{cases}
    \phi_{n0}(0)\,,
    \hspace*{4.35cm}\textrm{for contact interactions,}\\
    (1+\epsilon^2)^{-3/2}
    \Bigl(\frac{1-\epsilon^2}{1+\epsilon^2}\Bigr)^{n}\phi_{n0}(0)\,,
    \qquad\textrm{for smeared contact interactions.}
  \end{cases}
\end{equation}

As discussed in Sec.~\ref{sec:matrixelements}, the matrix
elements in the 2-body sector for $L=l=0$ are given by
\begin{equation}
  \label{eq:renorm2} 
  \bracket{n}{H^{\left(2\right)}}{n'}=
  \left(2n+\frac{3}{2}\right)\delta_{n,n'}+g^{(2)}f_{n\epsilon}f_{n'\epsilon}\:.
\end{equation} 
In the model space corresponding to $N$, the solution
$\left|\psi_{E^{(2)}}\right>$ of the eigenvalue problem with
the eigenvalue $E^{(2)}$ is expanded in oscillator functions
\begin{equation}
  \bket{\psi_{E^{(2)}}}=\sum_{n=0}^{N/2}c^{E^{(2)}}_n\bket{n}\:.
  \label{eq:basisexp}
\end{equation}
Applying the Hamiltonian to Eq.~\eqref{eq:basisexp} and
projecting on the oscillator state $\bket{k}$,
% \begin{align}
%   \bbracket{k}{H^{\left(2\right)}}{\psi_{E^{(2)}}}=
%   \sum_{n}^{N/2}c^{E^{(2)}}_n\bbracket{k}{H^{\left(2\right)}}{n}=
%   E^{(2)}\bbraket{k}{\psi_{E^{(2)}}}\;.
% \end{align}
% With equation \eqref{eq:renorm2} follows
we obtain
\begin{align}
  \label{eq:expansion}
  \left(2k+\frac{3}{2}\right)c^{E^{(2)}}_k+\sum_{n=0}^{N/2}g^{(2)}
  f_{k\epsilon} f_{n\epsilon} c^{E^{(2)}}_n &=E^{(2)}
  c^{E^{(2)}}_k\:.
\end{align}
% For $c^{E^{(2)}}_k$ follows
% \begin{align}
%   c^{E^{(2)}}_k&=g^{(2)}f_{k\epsilon}\sum_{n}^{N/2}\frac{f_{n\epsilon}c^{E^{(2)}}_n}{E^{(2)}-2k-\frac{3}{2}}\:.
% \end{align}
% Inserted in equation \eqref{eq:expansion} we find finally
% the renormalisation condition
Solving for $c^{E^{(2)}}_k$ and reinserting the result in
Eq.~\eqref{eq:expansion}, we find the running of the
coupling constant $g^{(2)}$ with $N$:
\begin{align}
  \label{eq:renorm_g2}
  \frac{1}{g^{(2)}(N)}&=
  -\sum_{n=0}^{N/2}\frac{f_{n\epsilon}^2}{2n+\frac{3}{2}-E^{(2)}}\:.
\end{align}

The three-body coupling constant $g^{(3)}$ is fixed by an
energy of the three-body system $E^{(3)}$. The solution
$\bket{\alpha_N}$ of the three-body system without
three-body interaction for a given value of $N$ is:
\begin{align}
  \bket{\alpha_N}=\sum\limits_{n_1l_1,n_2l_2,L}Z_N
  \left(\alpha_N;n_1l_1,n_2l_2,L\right)
  \bket{n_1l_1,n_2l_2,L}\:,
\end{align}
with energy eigenvalue $D_N(\alpha_N)$. Note that the
eigenstates are degenerate in the total angular momentum
projection $M$. With these states, we calculate the matrix
elements for the complete Hamiltonian including the
three-body interaction
\begin{multline}
  \bracket{\alpha_N}{H^{\left(3\right)}}{\alpha'_N}=
  D_N(\alpha_N)\delta_{\alpha_N,\alpha'_N}\\
  +g^{(3)}(N) \Bigl(\sum\limits_{n_1,n_2}
  Z_N\left(\alpha_N;n_10,n_20,0\right)
  f_{n_1\epsilon}f_{n_2\epsilon}\Bigr)
  \Bigl(\sum\limits_{n'_1,n'_2}
  Z_N\left(\alpha'_N;n'_10,n'_20,0\right)
  f_{n'_1\epsilon}f_{n'_2\epsilon}\Bigr)\:.
\end{multline}
% By comparison with Eq.~\eqref{eq:renorm2}
Requiring the energy of the state $\bket{\alpha_N}$ to be
$E^{(3)}$ for all $N$, the renormalization condition for the
three-body coupling follows:
\begin{equation}
  \label{eq:renorm_g3}
  \frac{1}{g^{(3)}(N)}= -\sum\limits_{\alpha_N}
  \frac{\left(\sum\limits_{n_1,n_2}
      Z_N\left(\alpha_N;n_10,n_20,0\right)
      f_{n_1\epsilon}f_{n_2\epsilon}\right)^2} 
  {D_N(\alpha_N)-E^{(3)}}\:. 
\end{equation}
Note that the expansion coefficients
$Z_N\left(\alpha_N;n_10,n_20,0\right)$ as well as the
eigenvalue $D_N(\alpha_N)$ explicitly depend on $N$.  The
full spectrum in a model space for given $N$ can then be
determined with these coupling constants by diagonalization
of the Hamiltonian matrix. Our general strategy for
calculating the Hamiltonian matrix for a system of $A$
identical bosons is described in
Appendix~\ref{sec:hamiltonmatrixdetails}.

\subsection{Running of $g^{(2)}$ and $g^{(3)}$}
\label{sec:running-g2-g3}

In this subsection, we study the running of the coupling
constants $g^{(2)}$ and $g^{(3)}$ for contact interactions
in detail.  Stetcu et al.~\cite{Stetcu:2010} rewrote the sum
in Eq.~\eqref{eq:renorm_g2} in terms of $\Gamma$-functions
and the generalized hypergeometric function $_3F_2$. They
found an explicit relation for $g^{(2)}(N)$ from which the
behavior for large $N$ can be obtained.

Here, we provide an alternative derivation of the behavior
of $g^{(2)}(N)$ for large values of $N$ using an integral
representation of the sum in Eq.~\eqref{eq:renorm_g2}.  For
$E^{(2)}<3/2$ each term in the sum is positive and the
denominator grows monotonously. We examine the behavior for
very large $N$. With the Euler-Maclaurin formula the sum can
be approximated as an integral
\begin{align}
  \label{eq:integralg2}
  -\frac{1}{g^{(2)}(N)}=
  \sum_{n=0}^{N/2}\frac{\phi_{n0}^2(0)}{2n+\frac{3}{2}-E^{(2)}}=
  \frac{1}{\pi^{3/2}} \int_{1}^{N/2} \frac{\Gamma(x+3/2)\;\d
    x} {\Gamma(3/2)\Gamma(x+1)
    (2x+\frac{3}{2}-E^{(2)})}+\mathcal{O}(1)\;,
\end{align}
where $\phi_{n0}(0)$ was substituted with
Eq.~\eqref{eq:phi0}. The quotient of Gamma functions can be
expanded in $x$,
\begin{align}
  \frac{\Gamma(x+3/2)}{\Gamma(3/2)\Gamma(x+1)}=
  \frac{1}{\Gamma(3/2)} \sqrt{x}+
  \mathcal{O}(\frac{1}{\sqrt{x}})\;.
\end{align}
Inserting this expansion in Eq.~(\eqref{eq:integralg2}) and
integrating, we find
\begin{align}
  g^{(2)}(N)&= -\frac{\pi^2}{\sqrt{2}}\frac{1}{\sqrt{N}}
  +\mathcal{O}(1/N)\;.
\end{align}
Thus the coupling constant vanishes as $1/\sqrt{N}$.
Identifying $\sqrt{N}$ with the momentum cutoff $\Lambda$,
this is consistent with the renormalization in free space
\cite{Bedaque:1998kg}.  We thus expect the leading errors
from finite $N$ in our effective theory to scale with
$1/\sqrt{N}$.

For $E^{(2)}<3/2$, the coupling rapidly approaches zero as
$N$ is increased.  In the case that $E^{(2)}>3/2$, the terms
in the sum in Eq.~\eqref{eq:integralg2} are negative at
first until $n>(E^{(2)}/2-3/4)$. The coupling $g^{(2)}(N)$
as a function of $N\in\mathbb{R}$ thus develops a minimum
for $N\rightarrow(E^{(2)}/2-3/4)$ and has a pole as $N$ is
increased further. For even larger $N$ it approaches zero as
well.

For smeared contact interactions an additional damping
factor appears. For Eq.~\eqref{eq:integralg2}, we have
\begin{align}
  -\frac{1}{g^{(2)}(N)}&=\frac{1}{(1+\epsilon^2)^{3}}
  \sum_{n=0}^{N/2}\frac{\phi_{n0}^2(0)}{(2n+\frac{3}{2}-E^{(2)})}
  \left(\frac{1-\epsilon^2}{1+\epsilon^2}\right)^{2n}\nonumber \\
  &= \frac{1}{\pi^{3/2}(1+\epsilon^2)^{3}}
  \int_{0}^{N/2}\frac{\Gamma(x+3/2)\,\d x}
  {\Gamma(3/2)\Gamma(x+1) (2x+\frac{3}{2}-E^{(2)})}
  \left(\frac{1-\epsilon^2}{1+\epsilon^2}\right)^{2x}
  +\mathcal{O}(1)\;.
\end{align}
The integral now converges to a constant for
$N\rightarrow\infty$ exponentially at a fixed value of
$\epsilon$. Thus, the coupling constant converges to a
finite number. Note, that the larger $\epsilon$ is, the
faster the coupling constant converges and the energy
spectrum becomes independent of $N$. This result reflects
the additional regularization of the contact interaction by
the smearing.

\begin{figure}[htbp]
  \centering
  \includegraphics[angle=270,
  width=0.6\linewidth]{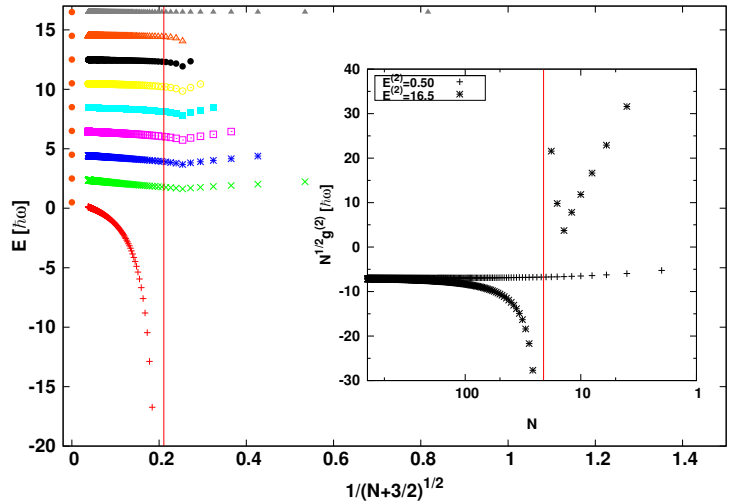}
  \caption{Energy spectrum in the two-body sector for
    $E^{(2)}=16.5$.  The dots on the left show the exact
    energies.  The inset shows the running coupling
    constants $\sqrt{N}\,g^{(2)}$ in the unitary limit for
    renormalization energies $E^{(2)}=0.5$ and
    $E^{(2)}=16.5$.}
  \label{fig:couplingg2}
\end{figure}

In the inset of Fig.~\ref{fig:couplingg2}, we illustrate the
behavior of the running coupling constant $\sqrt{N} g^{(2)}$
for $\epsilon=0$. We show $\sqrt{N}\,g^{(2)}$ in the unitary
limit for the renormalization energies $E^{(2)}=0.5$ (ground
state) and $E^{(2)}=16.5$ (8th excited state). In the case
of $E^{(2)}=0.5$, the running coupling $\sqrt{N}\,g^{(2)}$
has already converged to a constant value for small values
of $N$ and shows no structure.  For $E^{(2)}=16.5$, the
situation is as described above. The coupling constant
approaches zero for $N$ around 14 where the denominator of
the right hand side of Eq.~\eqref{eq:integralg2} changes
sign.  Moreover, it changes from large positive to large
negative values around $N=21$.  This behavior can be
understood by looking at the spectrum for $E^{(2)}=16.5$
shown in Fig.~\ref{fig:couplingg2}. For small values of $N$,
the model space is not large enough to describe the ground
state adequately. The behavior of the coupling constant
$\sqrt{N}\,g^{(2)}$ is exactly such that this new state
enters the renormalized spectrum from minus infinity keeping
the other states unchanged.  In the continuum case, a
similar behavior is observed for the three-body spectrum
\cite{Bedaque:1998kg,HammerRMP2012}. For larger values of
$N$, when the model space is large enough to describe the
ground state, the running coupling $\sqrt{N}\,g^{(2)}$
approaches the same value for both cases.
%%%%%%%%%%%%%%%%%%%%%%%%%%%%%%%%%%%%%%%%%%%%%%%%%%%%%%%%%%%%
%%%%%%%%%%%%%%%%%%%%%%%%%%%%%%%%%%%%%%%%%%%%%%%%%%%%%%%%%%%%

For the coupling constant $g^{(3)}$ the situation is more
complicated. The eigenvalues $D(\alpha)$ as well as the
eigenstates now depend on the cutoff parameter $N$.
Therefore it is not straightforward to derive the
leading-order behavior of $g^{(3)}$ analytically.  Werner et
al.\ have provided a semi-analytic solution for the
three-body problem in a harmonic trap in the unitary
limit~\cite{Werner:2006}.  We use their results for the
energy spectrum to benchmark our three-body calculations.

\begin{figure}[htbp]
  \centerline{
    \includegraphics[angle=270,
    width=0.5\linewidth]{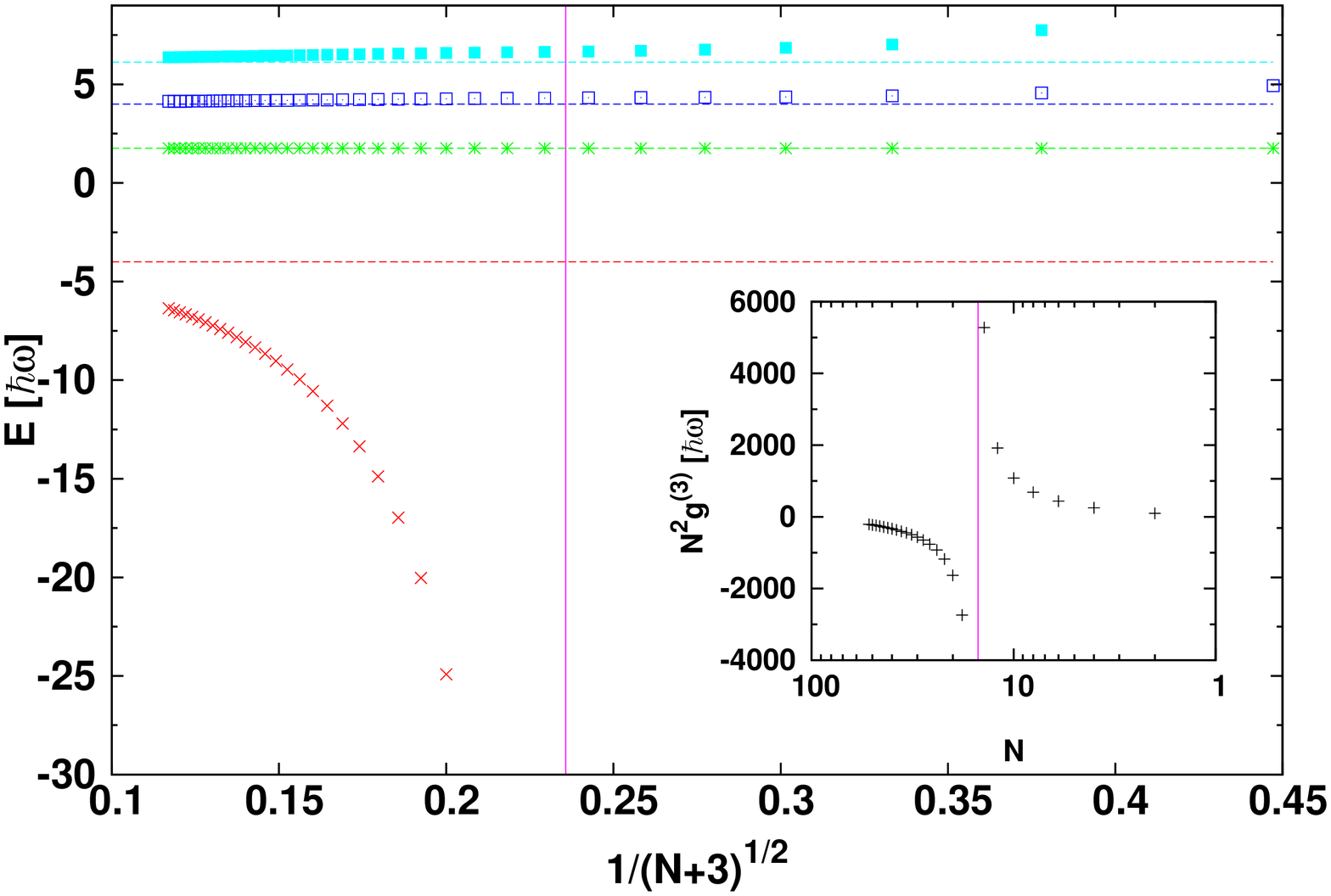}
    \quad \includegraphics[angle=270,
    width=0.5\linewidth]{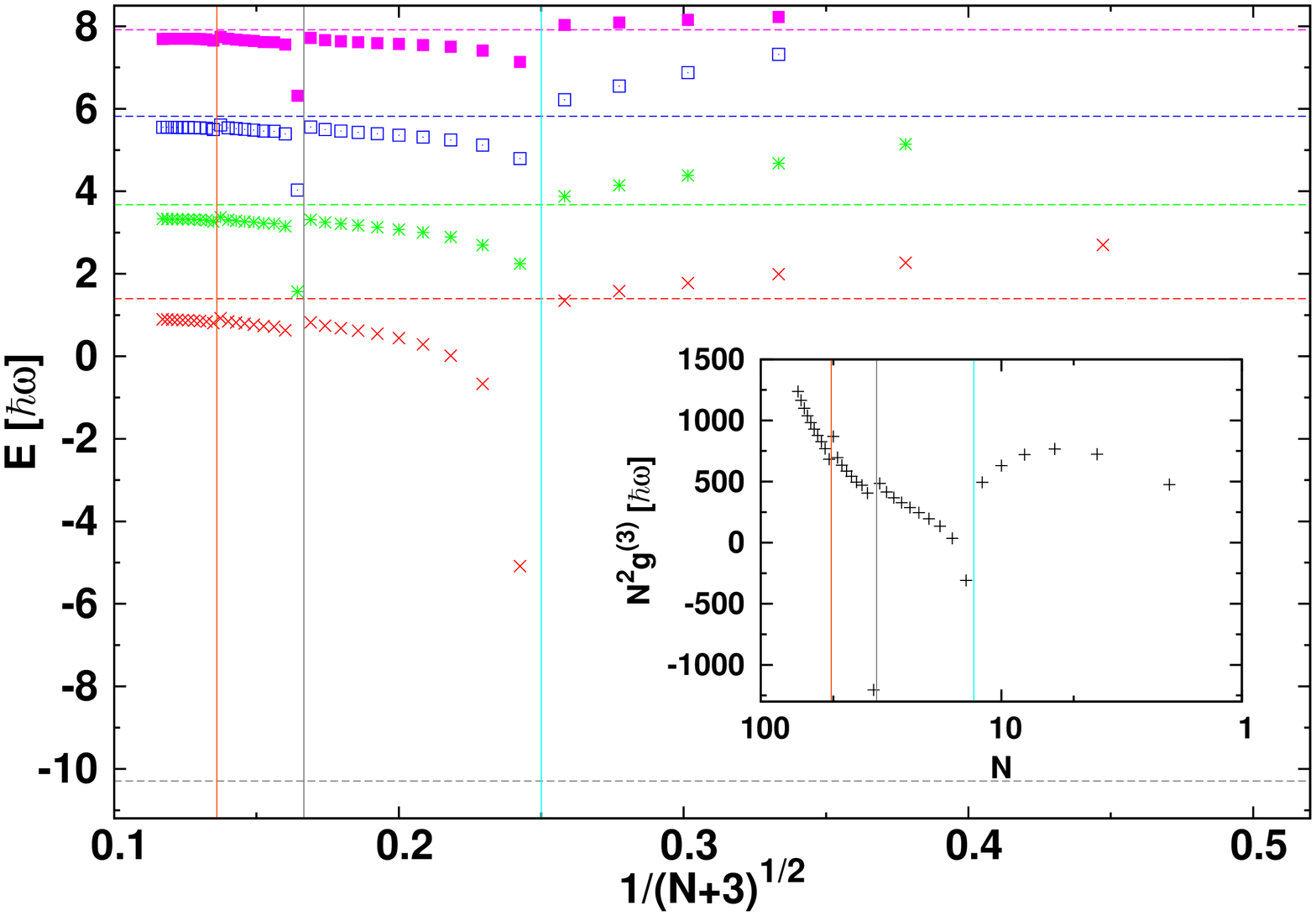}}
  \caption{Left panel: Energy spectrum of Efimov-like states
    in the three-boson sector in the unitary limit
    ($E^{(2)}=0.5$) for $E^{(3)}\approx1.76$.  The dashed
    lines indicate the exact energies \cite{Werner:2006}.
    Right panel: Energy spectrum for Efimov-like states in
    the unitary limit $E^{(2)}=0.5$ with
    $E^{(3)}\approx16.14$ for contact interactions. The
    running coupling constants $N^2\,g^{(3)}$ corresponding
    to both cases are shown in the insets. Poles are
    indicated by vertical solid lines.}
  \label{fig:couplingg3_1.76}
\end{figure}

In the left panel of Fig.~\ref{fig:couplingg3_1.76}, we show
the spectrum in the unitary limit for a three-body
renormalization energy $E^{(3)}\approx1.76$ as a function of
$N$. At $N=16$ a new three-body state enters the spectrum of
the model space from minus infinity.  For large $N$ this
state approaches the exact eigenenergy at $-4$. In the
inset, the corresponding coupling constant $g^{(3)}$ is
shown. As in the two-body case, the coupling constant
diverges, here around $N=16$, and changes from large
positive to large negative values. The three-body spectrum
and the coupling constant in the continuum case show the
same behavior \cite{Bedaque:1998kg}. The phenomena are the
same as in the two-body case discussed above.
%%%%%%%%%%%%%%%%%%%%%%%%%%%%%%%%%%%%%%%%%%%%%%%%%%%%%%%%%%%%
%%%%%%%%%%%%%%%%%%%%%%%%%%%%%%%%%%%%%%%%%%%%%%%%%%%%%%%%%%%%
In the right panel of Fig.~\ref{fig:couplingg3_1.76}, we
present the energy spectra for the renormalization
$E^{(3)}\approx16.14$. This renormalization belongs to the
energy spectrum in the full Hilbert space: $\cdots$,
$-5319.28$, $-10.29$, $1.4$, $3.68$, $5.82$, $7.92$, $9.99$,
$12.05$, $14.10$, $16.14$, $\cdots$.  The states with
energies lower than $1.4$ cannot be found inside the
feasible model spaces for this renormalization
energy. Remarkably, the states approach the exact energies
but at specific values of the cutoff marked by vertical
lines the spectrum rearranges.  The smallest value of $N$
corresponds to the appearance of a ground state at energy
$1.4$.

At the two largest cutoff values marked, there are small
discontinuities in the energy. These discontinuities are
artifacts of the renormalization method and do not
correspond to new states entering the spectrum. Since we
work in a finite model space, not only the Efimov-like
states but also the universal states depend very weakly on
the three-body interaction. In the special case when for a
given cutoff $\bar{N}$ the renormalization energy $E^{(3)}$
exactly coincides with the eigenvalue of a universal state
corresponding to another spectrum with renormalization
energy $\bar{E}^{(3)}$, our computer code erroneously
renormalizes to the spectrum characterized by
$\bar{E}^{(3)}$. The energy values in the neighborhood of
such points should therefore be discarded.

In the inset of the right panel of
Fig.~\ref{fig:couplingg3_1.76}, the behavior of the
corresponding coupling constant is shown. The coupling
constant has a salient behavior at three positions indicated
by the vertical lines.  At the smallest value of $N$ where
the ground state enters into the spectrum from below, the
coupling changes from large positive to large negative
values.  The approximate pole which is expected at this
position due to the discrete nature of the cutoff $N$ is
strongly distorted on the right wing due to finite cutoff
effects. At the two larger values of $N$, there are
discontinuities due to the renormalization artifacts
discussed above.  Moreover, for even larger $N$ the coupling
approaches an approximate pole corresponding to the addition
of a new ground state at energy $-10.29$. This pole is not
reached in our calculation; it would require a larger $N$.

\subsection{Error Analysis}
\label{sec:corr-due-scal}

There are various sources of errors in our calculation.  In
this subsection, we perform a detailed error analysis.

\begin{figure}[htbp]
  \centering
  \includegraphics[angle=270,width=0.6\linewidth]{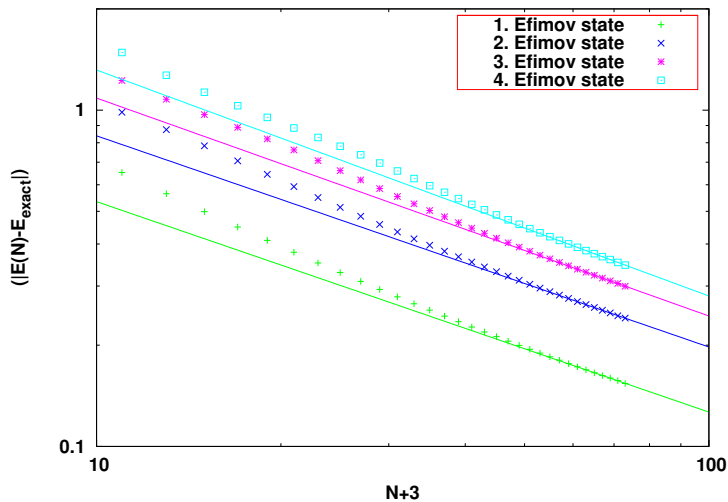}
  \caption{Corrections in energy eigenvalues as a function
    of $N+3$ with $A=3$, $E^{(2)}=0.5$, $E^{(3)}=-1$.}
  \label{fig:corr-due-to-N}
\end{figure}

First, there are corrections due to the ultraviolet cutoff
parameter $N$.  Our considerations in
subsection~\ref{sec:running-g2-g3} showed that the finite
model space also implies an infrared cutoff which vanishes
as $N$ is increased.  The errors due to both cutoffs show
the same scaling behavior in $N$. Therefore, we expect
corrections in the energy eigenvalues of order
$1/\sqrt{N+(A-1)3/2}$ for large $N$.  The shift of $N$ by
$(A-1)3/2$ under the square root takes into account the
zero-point energy of the free $A$-body system.  We can
extract the scaling behavior of these corrections from the
error analysis introduced by Lepage \cite{Lepage:1997cs}.
In Fig.~\ref{fig:corr-due-to-N}, we show the deviation of
the lowest three-body energy eigenvalues from the exact
values in the unitary limit $E^{(2)}=0.5$ with three-body
renormalization energy $E^{(3)}=-1$. The double-logarithmic
plot shows a linear dependence of the energy differences on
$\log(N+3)$ for large $N$ values. From a linear fit to the
five largest values of $N$, we find a slope of $s\approx
-0.6$ for the first four Efimov states above the state used
for renormalization.  The small difference to the expected
value of $s=-0.5$ could be due to contamination from higher
order corrections. As a consequence, the cutoff dependence
for contact interactions is in agreement with power counting
arguments based on identifying the continuum cutoff
$\Lambda$ with $1/\sqrt{N+(A-1)3/2}$. A similar power law
dependence of the leading corrections to three-body energies
for contact interactions was observed by Furnstahl et
al.~\cite{Furnstahl:2012private}.
%%%%%%%%%%%%%%%%%%%%%%%%%%%%%%%%%%%%%%%%%%%%%%%%%%%%%%%%%%%%
%%%%%%%%%%%%%%%%%%%%%%%%%%%%%%%%%%%%%%%%%%%%%%%%%%%%%%%%%%%%
% Remarkably, the cutoff dependence show no exponential
% decrease predicted for finite range potentials by
% Furnstahl et.~al.~\cite{Furnstahl:2012}.

In the case of smeared contact interactions, we find an
exponential dependence on $\sqrt{N}$. In the two-body
sector, we consider model spaces with cutoffs up to
$N=700$. In a Lepage plot of $\log(|E_N-E_{\infty}|)$ we
find for large cutoffs a slope of about $-0.5$. This implies
that for large $N$ the calculated energies behave as
\begin{align}
  \label{eq:2}
  E_N\approx E_{\infty}+c_1e^{-c_2\sqrt{N}}\;.
\end{align}
Thus the smearing changes the leading corrections from an
inverse power law to an exponential behavior.

The convergence properties of variational calculations using
basis expansions have been studied already in the 1970's
\cite{Delves:1972,Schneider:1972}.  More recently, the
convergence properties of ab initio calculations of light
nuclei in a harmonic oscillator basis were
investigated~\cite{Coon:2012,Furnstahl:2012}.  Due to their
singularity, we expect the contact interactions to behave
quite differently, but it is interesting to compare our
results for smeared interactions to
Refs.~\cite{Coon:2012,Furnstahl:2012}.  However, one has to
keep in mind that we consider a physical trapping potential
instead of a mere basis expansion.  Moreover, in our
calculation the effective interaction is renormalized at
each $N$ in order to keep the energies of a given two- and
three-body state fixed.

Our results show the same exponential dependence on the
ultraviolet cutoff and on the infrared cutoff
$\lambda_{sc}\propto 1/\sqrt{N+3/2}$ as observed by Coon et
al.\ for the Idaho N3LO potential~\cite{Coon:2012}.
Furnstahl et al.~\cite{Furnstahl:2012}, in contrast,
observed a Gaussian dependence on the UV cutoff
\begin{align}
  \label{eq:2x}
  E_N\approx E_{\infty}+A_0 e^{-A_1 (\sqrt{N})^2}\;,
\end{align}
using similarity renormalization group (SRG)-evolved chiral
interactions, but their dependence on the infrared cutoff is
consistent with ours.  The difference might be due to the
use of SRG-evolved interactions, but this question requires
further study.

If we want to extract the energies for genuine contact
interactions from smeared contact interactions, there are
also errors due to the width parameter $\epsilon$ which
generates a finite interaction range. The corresponding
two-body problem in free space is separable and can be
solved analytically.  In our calculations, we determine the
coupling constants from matching to energy levels in the
oscillator.  One two-body energy level $E^{(2)}$ is kept
constant in each model space and the coupling constant
$g^{(2)}$ is determined from this matching condition
(cf.~Eq.~\eqref{eq:renorm_g2}). We find:
\begin{align}
  \label{eq:8}
  \frac{1}{g^{(2)}_N(\epsilon)}&=
  -\Bigl(\bigl(1+\epsilon^2\bigr)\sqrt{\pi}\Bigr)^{-3}
  \sum_{n=0}^{N/2}
  \frac{\Bigl(\frac{1-\epsilon^2}{1+\epsilon^2}\Bigr)^{2n}
    \frac{(2n+1)!!}{n!\,2^{n}}}
  {2n+\frac{3}{2}-E^{(2)}}\:,\nonumber \\
  &\overset{N\rightarrow\infty}{\longrightarrow}
  \Bigl(\bigl(1+\epsilon^2\bigr)\sqrt{\pi}\Bigr)^{-3}
  \bigl(2E^{(2)}-3\bigr)^{-1}\;
  _2\F_1\Bigl(\frac{3}{2},\frac{3}{4}-\frac{E^{(2)}}{2};
  \frac{7}{4}-\frac{E^{(2)}}{2};
  \frac{(\epsilon^2-1)^2}{(\epsilon^2+1)^2}\Bigr)\,,
\end{align}
where $_2\F_1(a,b;c;z)$ denotes a hypergeometric function.
The limit $N\rightarrow\infty$ yields the coupling constant
in the full Hilbert space.

We use $\epsilon$ as a smearing parameter and are interested
in the behavior for small $\epsilon$. Expanding the first
and second line of Eq.~\eqref{eq:8}, we find:
\begin{align}
  \label{eq:5}
  g^{(2)}_{\infty}(\epsilon)&=-2\pi^{\frac{3}{2}}\,\epsilon-
  \frac{4\pi^2\Gamma(\frac{3}{4}-\frac{E^{(2)}}{2})}
  {\Gamma(\frac{1}{4}-\frac{E^{(2)}}{2})}\,\epsilon^2-
  2\pi^{\frac{3}{2}}\Bigl(1+4E^{(2)}+
  \frac{4\pi\Gamma(\frac{3}{4}-\frac{E^{(2)}}{2})^2}
  {\Gamma(\frac{1}{4}-\frac{E^{(2)}}{2})^2}\Bigr)\,\epsilon^3+
  \mathcal{O}(\epsilon^4)\,,\\
  g^{(2)}_{N}(\epsilon)&=c_0(N,E^{(2)})+c_2(N,E^{(2)})\epsilon^2
  +\mathcal{O}(\epsilon^4)\,.
\end{align}
Remarkably, the results for finite and infinite $N$ differ
fundamentally since the limits $\epsilon\to 0$ and
$N\to\infty$ do not commute. For $N\rightarrow\infty$, the
coupling constant depends linearly on $\epsilon$ in leading
order and vanishes for $\epsilon\rightarrow 0$. For finite
$N$, the leading term is independent of $\epsilon$ and the
first correction is of order $\epsilon^2$. The constant term
in $\epsilon$ is a consequence of regularization with $N$
but the vanishing of the correction linear in $\epsilon$ is
unexpected.  The expansion parameter $c_0(N,E^{(2)})$
vanishes like $1/\sqrt{N}$ for $N\rightarrow\infty$.

\begin{figure}[htbp]
  \centering
  \includegraphics[angle=0,width=0.4\linewidth]{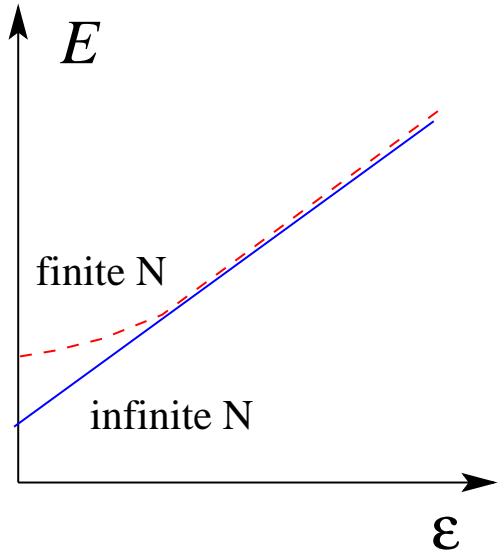}
  \caption{Schematic dependence of energies on the smearing
    parameter $\epsilon$ for finite $N$ (dashed line) and
    infinite $N$ (solid line).}
  \label{fig:extra-N}
\end{figure}

The corresponding two-body energies show a similar behavior.
First, the value of $g^{(2)}$ is fixed for a given
$\epsilon$ and $E^{(2)}$. The other energies $E^{(2)}_i$ can
then be calculated numerically using a root-finding
algorithm.  For $N=\infty$ and small $\epsilon$ the energies
$E^{(2)}_i(\epsilon,E^{(2)})$ depend linearly on
$\epsilon$. In order to extract the results for a zero-range
contact interaction from a calculation with smeared
interactions, a linear extrapolation in $\epsilon$ is thus
appropriate.  For finite $N$, however, the dependence of the
excitation energies $E^{(2)}_i(\epsilon,E^{(2)},N)$ on
$\epsilon$ is different. For $\epsilon\rightarrow 0$, we
find
$|E^{(2)}_i(\epsilon,E^{(2)},N)-E^{(2)}_i(0,E^{(2)},N)|\propto\epsilon^2$
and an extrapolation in $\epsilon$ does not reproduce the
result for a zero-range contact interaction.  The
$\epsilon$-dependence of the energies is illustrated
schematically in Fig.~\ref{fig:extra-N}.
%%%%%%%%%%%%%%%%%%%%%%%%%%%%%%%%%%%%%%%%%%%%%%%%%%%%%%%%%%%%
%%%%%%%%%%%%%%%%%%%%%%%%%%%%%%%%%%%%%%%%%%%%%%%%%%%%%%%%%%%%
In order to extract the contact interaction results from
smeared interactions, it is therefore important to
extrapolate in $N$ first (or use numerically converged
values) and then extrapolate to $\epsilon=0$ in a second
step.

The variation of the energies with $\epsilon$ can be
interpreted as a dependence on the effective range $r_e$ of
the two-body interaction. The effective range parameters in
terms of the effective coupling constant $g^{(2)}$ in the
continuum case are (See Appendix~\ref{sec:ERE_SGI}):
\begin{align}
  \label{eq:1}
  \frac{1}{a}&= \frac{\sqrt{2}\pi}{g^{(2)}}
  +\frac{1}{\sqrt{2\pi}\epsilon} \qquad \mbox{ and }\qquad
  r_e =\frac{\sqrt{2}\epsilon}{\sqrt{\pi}}-
  \frac{\sqrt{8}\pi\epsilon^2}{g^{(2)}}\,.
\end{align}
These relations are valid in the full Hilbert space but not
in the restricted model space characterized by finite $N$.
Analog to the case without a trap, our numerical results and
Eqs.~\eqref{eq:1} then imply that the leading corrections to
the energies $E^{(2)}_i$ for contact interactions are linear
in the effective range $r_e$.

\begin{figure}[htbp]
  \centerline{
    \includegraphics[angle=270,width=0.5\linewidth]{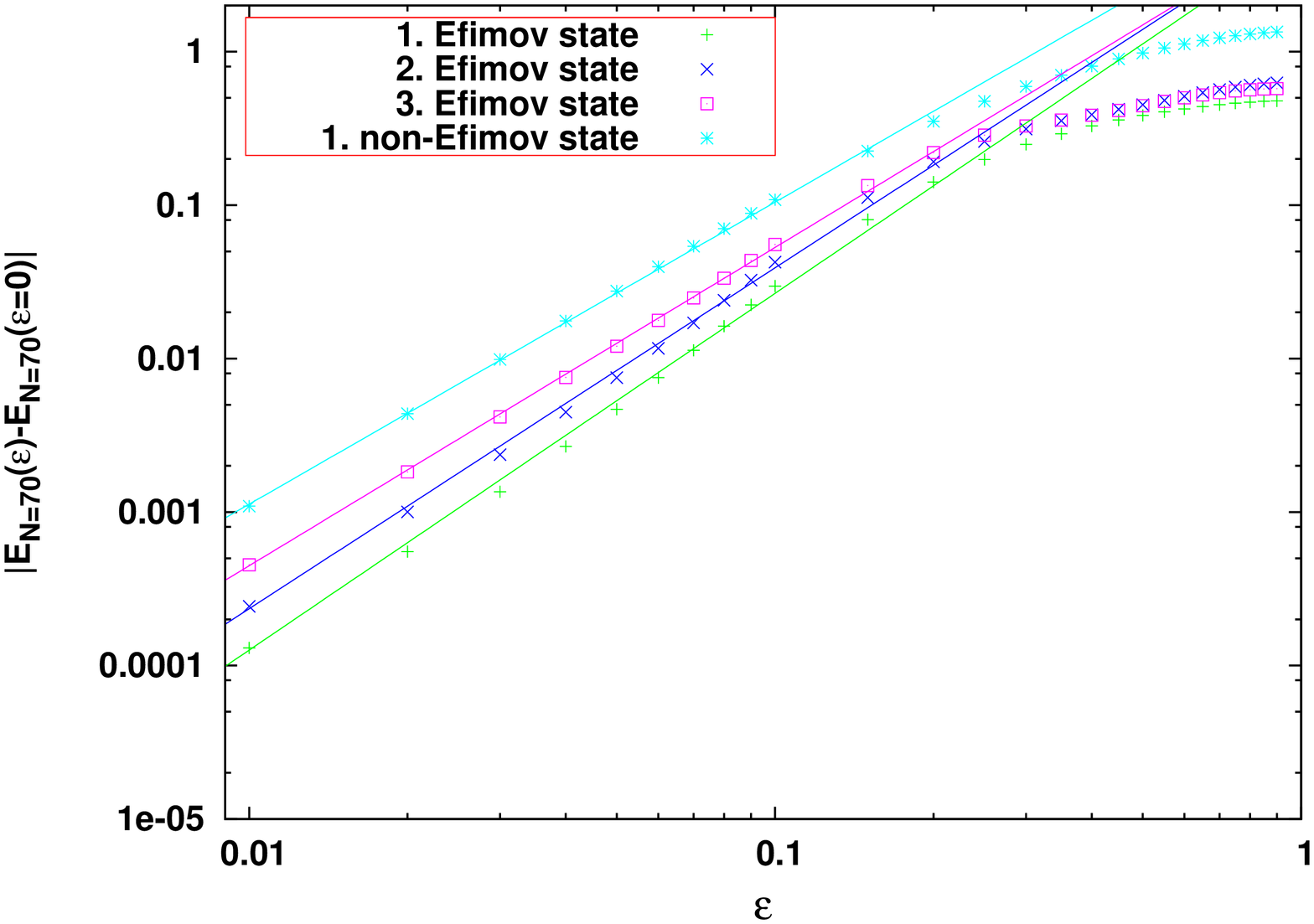}
    \quad
    \includegraphics[angle=270,width=0.5\linewidth]{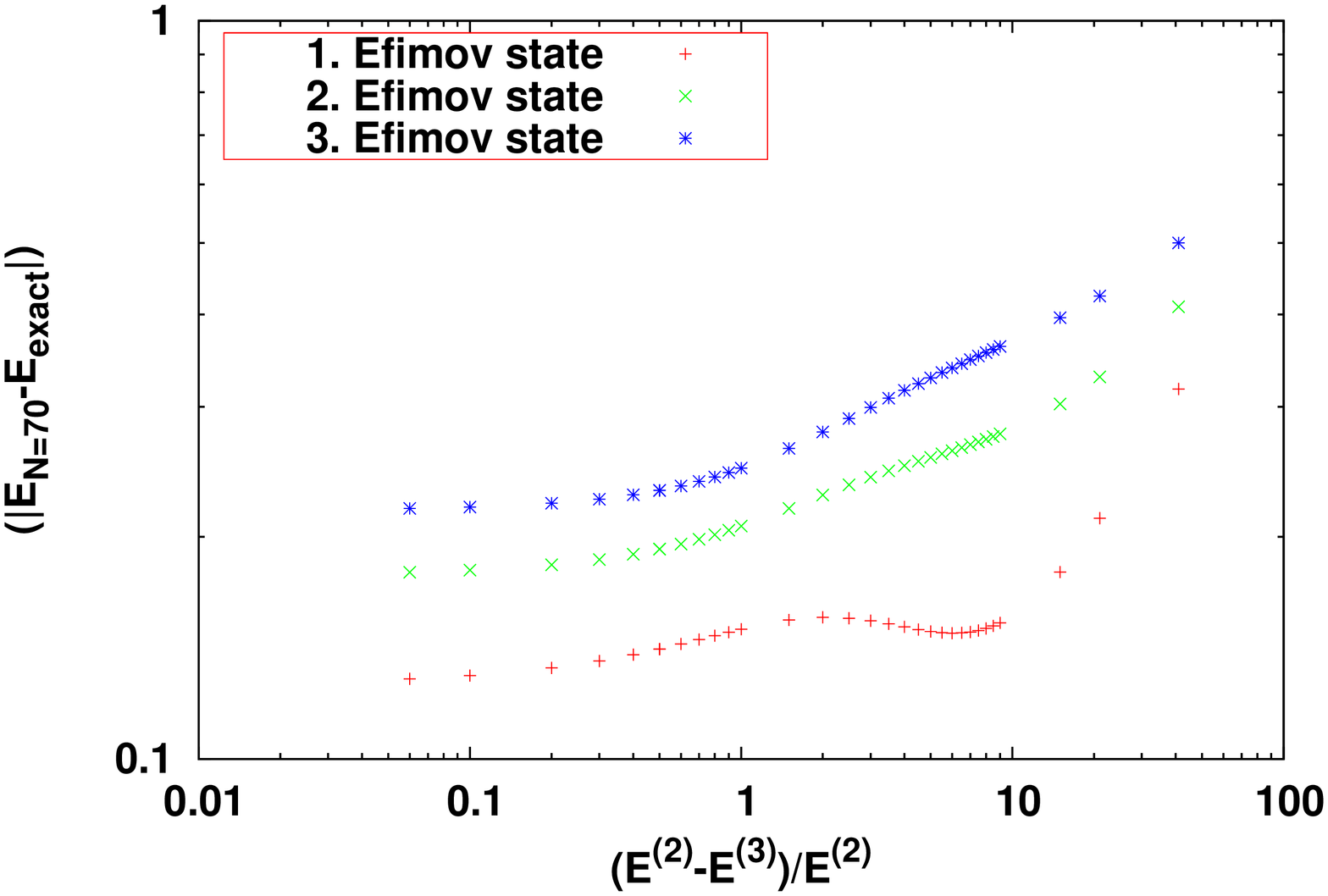}}
  \caption{Left panel: Corrections in energy eigenvalues as
    a function of $\epsilon$ with $A=3$, $E^{(2)}=0.5$,
    $E^{(3)}=0.5$.  Right panel: Corrections in
    $(E^{(2)}-E^{(3)})/E^{(2)}$ for the unitary limit with
    $A=3$.}
  \label{fig:corr-due-to-epsilon}
\end{figure}

We have also studied the $\epsilon$-dependence in the
three-body sector.  In the left panel of
Fig.~\ref{fig:corr-due-to-epsilon}, we show the corrections
from finite $\epsilon$ for finite $N$.  We plot the
deviation of results for smeared interactions with
$\epsilon\neq0$ from contact-interactions results with
$N=70$. We pick out the unitary limit for contact
interactions, which corresponds to $E^{(2)}=0.5$. Please
note, that we keep the two-body ground state constant while
$\epsilon$ is changed. For finite $\epsilon$, the system is
not exactly in the unitary limit anymore. For small
$\epsilon$ there is a linear dependence in the
double-logarithmic plot. The fit yields a slope round about
$s=2$ for the state presented, which implies a quadratic
dependency of the corrections in $\epsilon$ for
$\epsilon\rightarrow0$ in agreement with the behavior in the
2-body sector.
%%%%%%%%%%%%%%%%%%%%%%%%%%%%%%%%%%%%%%%%%%%%%%%%%%%%%%%%%%%%
%%%%%%%%%%%%%%%%%%%%%%%%%%%%%%%%%%%%%%%%%%%%%%%%%%%%%%%%%%%%

Finally, there are corrections from a mismatch in
renormalization energies $E^{(2)}$ and $E^{(3)}$. In an
effective theory, only states with excitation energies small
compared to the cutoff scale of the effective theory can be
described. Clearly, the renormalization energies must be
chosen in this energy range.  For $E^{(2)}\neq E^{(3)}$, we
expect errors governed by $(E^{(2)}-E^{(3)})$ from the
mismatch in the renormalization energies.  For contact
interactions ($\epsilon=0$), we have investigated the errors
in $(E^{(2)}-E^{(3)})$ numerically.  For this purpose, we
chose the unitary limit and varied $E^{(3)}$. The right
panel of Fig.~\ref{fig:corr-due-to-epsilon} shows the
deviation of the exact results from results in the model
space for various $(E^{(2)}-E^{(3)})$. As expected, the
corrections grow with increasing $|E^{(2)}-E^{(3)}|$. Thus
in practice, it is desirable to choose $E^{(2)} \approx
E^{(3)}$ to minimize this type of errors.

\section{Results}
We will now use these insights to study the energy spectra
of $A$-boson systems for $A=3,4,5,6$ in a trap. We will
follow two strategies to obtain the spectra from numerical
calculations:
\begin{enumerate}
\item use contact interactions and extrapolate in
  $1/\sqrt{N+(A-1)3/2}$ as in our previous work
  \cite{toelle2010}, and
\item use smeared contact interactions and extrapolate the
  converged results in $N$ linearly in the width parameter
  $\epsilon$.
\end{enumerate}
The difference between the two methods will be used to
estimate the errors in our calculation.

\subsection{Three identical bosons}

\begin{figure}[htbp]
  \centerline{
    \includegraphics[angle=270,
    width=0.5\linewidth]{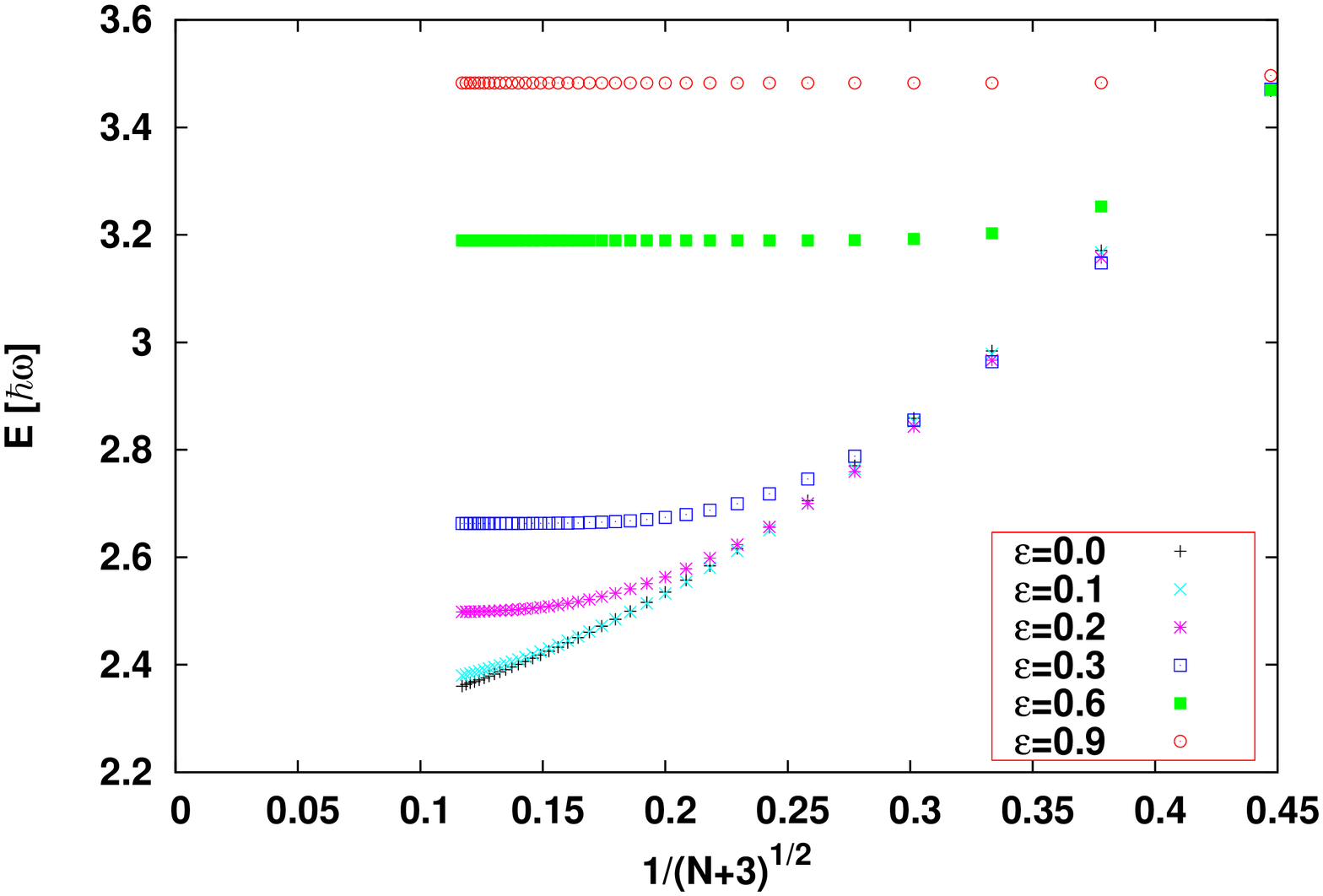}
    \quad
    \includegraphics[angle=270,
    width=0.5\linewidth]{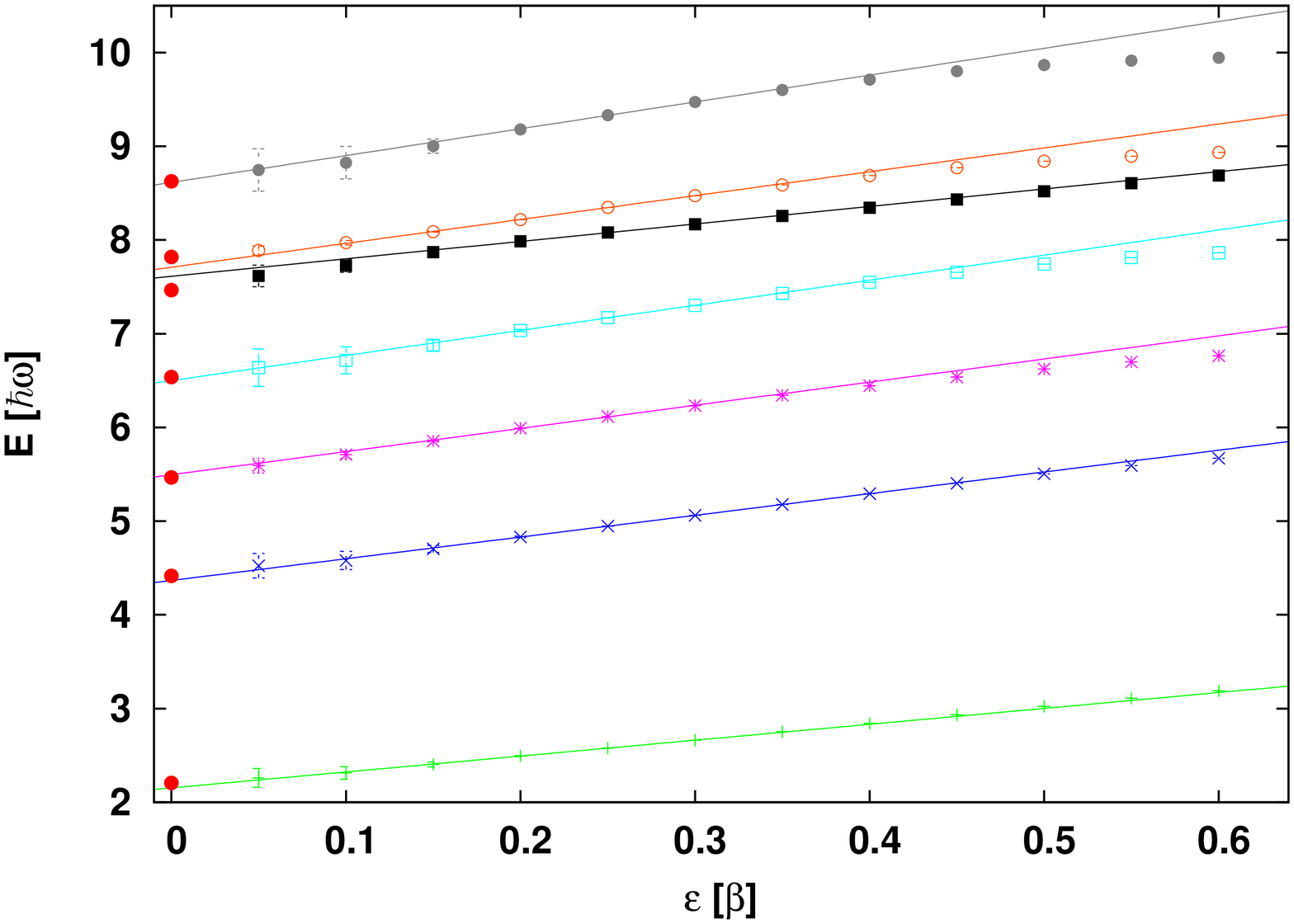}
  }
  \caption{Left panel: Eigenvalue for the first excited
    $L^{\pi}=0^+$ state with $A=3$ as a function of $N$ with
    $E^{(2)}=0.5$ and $E^{(3)}=-1$ for various $\epsilon$.
    Right panel: The converged eigenvalues as a function of
    $\epsilon$. $L^{\pi}=0^+$, $E^{(2)}=0.5$ and
    $E^{(3)}=-1$. The points at $\epsilon=0$ denote the
    exact eigenvalues from \cite{Werner:2006}. The dashed
    lines are linear fits to the values for $\epsilon\leq
    0.3$ used to extrapolate to $\epsilon=0$.  }
  \label{fig:gauss3B}
\end{figure}

At first, we consider the 3-boson sector. In the left panel
of Fig.~\ref{fig:gauss3B}, the eigenvalues of the first
excited $L^{\pi}=0^+$ state are shown as a function of the
cutoff parameter $N\leq70$ for various smearing parameters
$\epsilon$ in oscillator lengths $\beta$. The corresponding
renormalization energies are $E^{(2)}=0.5$ and $E^{(3)}=-1$.
%%%%%%%%%%%%%%%%%%%%%%%%%%%%%%%%%%%%%%%%%%%%%%%%%%%%%%%%%%%%%%%%%%%%
%%%%%%%%%%%%%%%%%%%%%%%%%%%%%%%%%%%%%%%%%%%%%%%%%%%%%%%%%%%%%%%%%%%%%%%%
For small model spaces the results are identical for all
interaction widths. With increasing interaction width
$\epsilon$ the eigenvalues converge for smaller values of
N. The converged eigenvalues do not coincide with the
results for contact interactions. However, as discussed in
the previous section, the parameter $\epsilon$ can now be
used as a extrapolation parameter instead of $N$.

The converged eigenvalues are given as a function of
$\epsilon$ in the right panel of Fig.~\ref{fig:gauss3B} for
the 5 lowest $L^{\pi}=0^+$ excited states. The solid lines
are linear extrapolations of the data points for
$\epsilon\leq 0.3$ and the results agree with the exact
values known for contact interactions in the unitary limit
\cite{Werner:2006} within $3\%$ errors.  The error bars of
the data points are estimated by the difference of the
eigenvalues related to the two highest cutoff parameters
$N$, i.e.\ with these two eigenvalues the exact one is
approximated by linear extrapolation and the error is
estimated as the difference between the state with the
highest cutoff parameter and the estimated one. In
Fig.~\ref{fig:gauss3B} the two model space sizes correspond
to $N=70$ and $N=68$.

\begin{figure}[htbp]
  \centering
  \includegraphics[angle=270,
  width=0.49\linewidth]{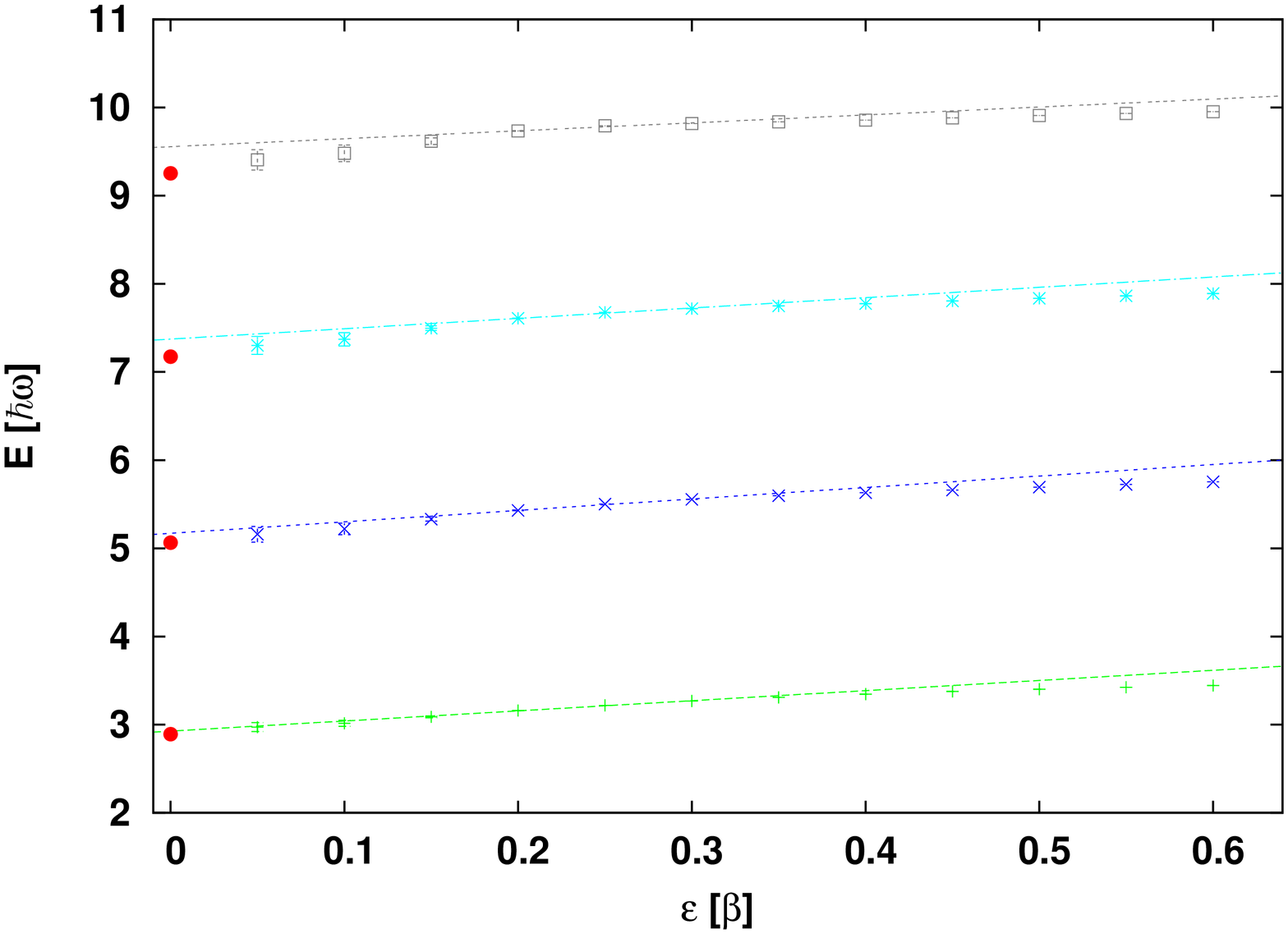}
  \includegraphics[angle=270,
  width=0.49\linewidth]{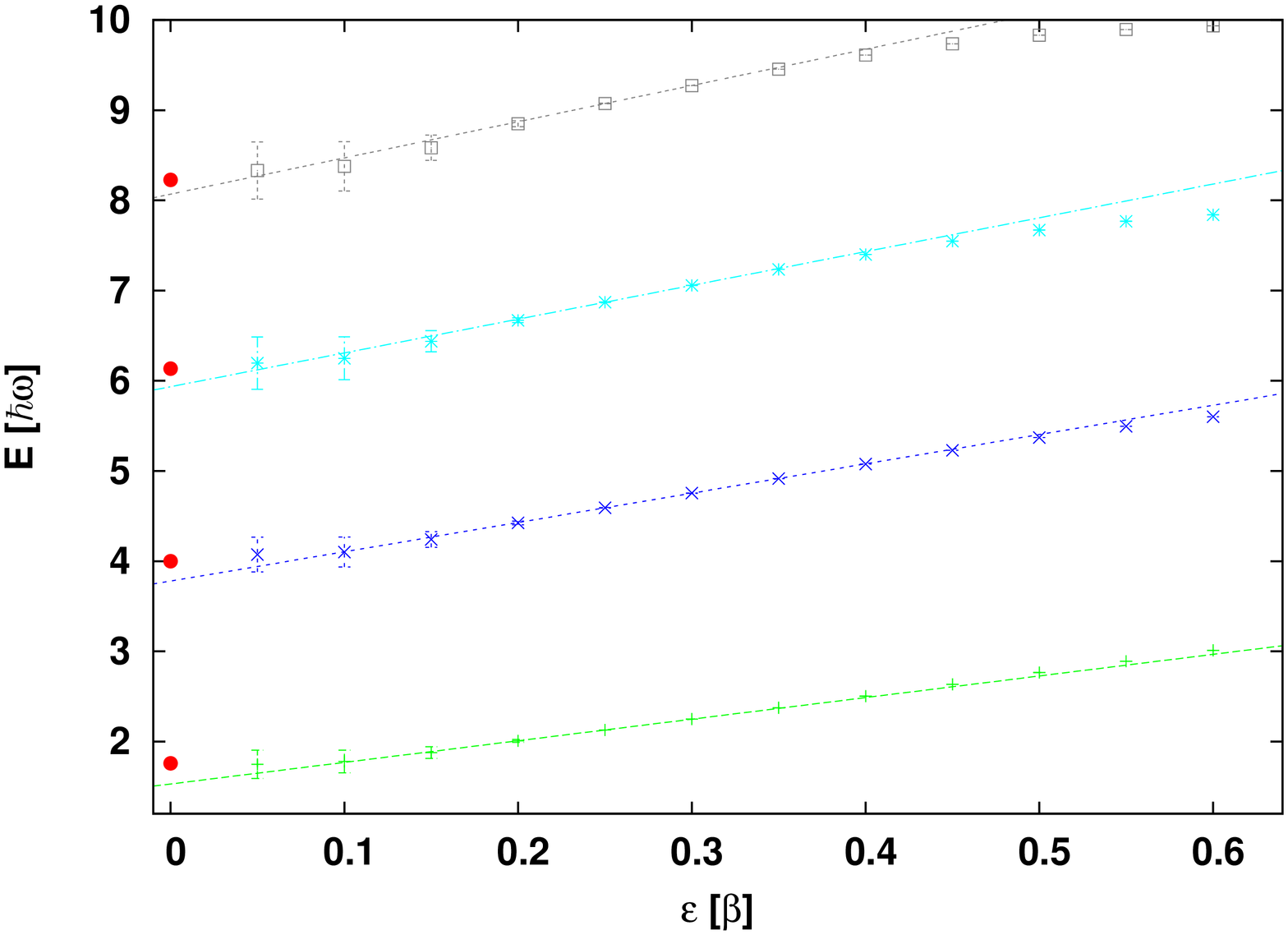}
  \caption{Spectrum of Efimov-like $L^{\pi}=0^+$ states with
    $A=3$ in the unitary limit as a function of
    $\epsilon$. At $\epsilon=0$ exact eigenenergies are
    drawn in. Left panel: Renormalization energy
    $E^{(3)}=0.5$. Right panel: Renormalization energy
    $E^{(3)}=-4$.}
  \label{fig:3BGauss0.5-4}
\end{figure}

In Fig.~\ref{fig:3BGauss0.5-4}, we show the spectra of
Efimov-like states in the unitary limit ($E^{(2)}=0.5$) for
$E^{(3)}=0.5$ and $E^{(3)}=-4$ in order to illustrate the
significance of the mismatch in renormalization scales. With
a linear extrapolation the exact values for $E^{(3)}=-4$ are
undershot systematically.  For larger $E^{(3)}$ the
extrapolation in $N$ achieves better results than a linear
extrapolation in $\epsilon$.
%%%%%%%%%%%%%%%%%%%%%%%%%%%%%%%%%%%%%%%%%%%%%%%%%%%%%%%%%%%%%%%%%%
%%%%%%%%%%%%%%%%%%%%%%%%%%%%%%%%%%%%%%%%%%%%%%%%%%%%%%%%%%%%%%%%%%%%
Our method clearly has larger errors for $E^{(3)}=-4$.

\begin{figure}[ht]
  \centering
  \includegraphics[angle=270,
  width=0.49\linewidth]{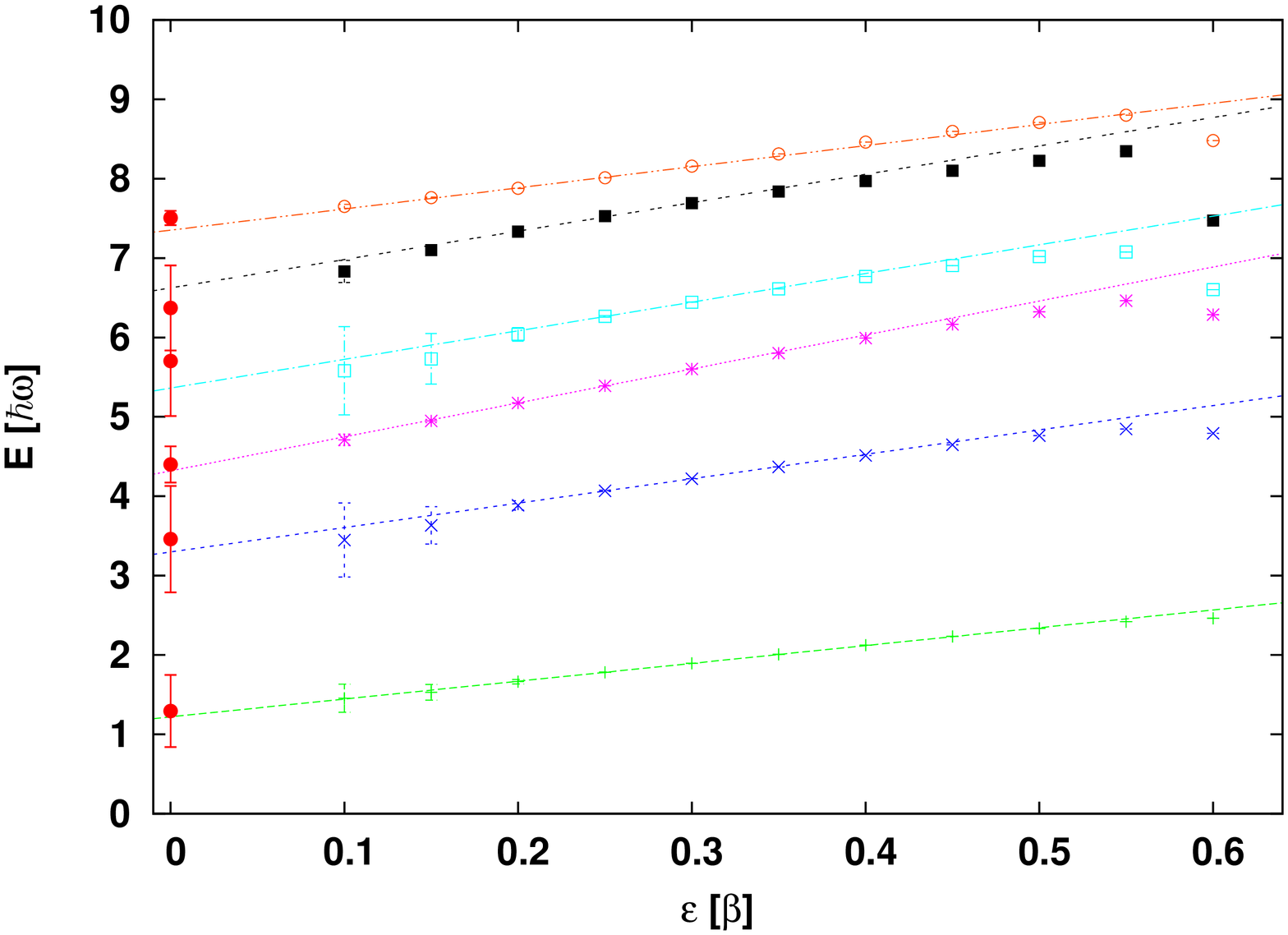}
  \includegraphics[angle=270,
  width=0.49\linewidth]{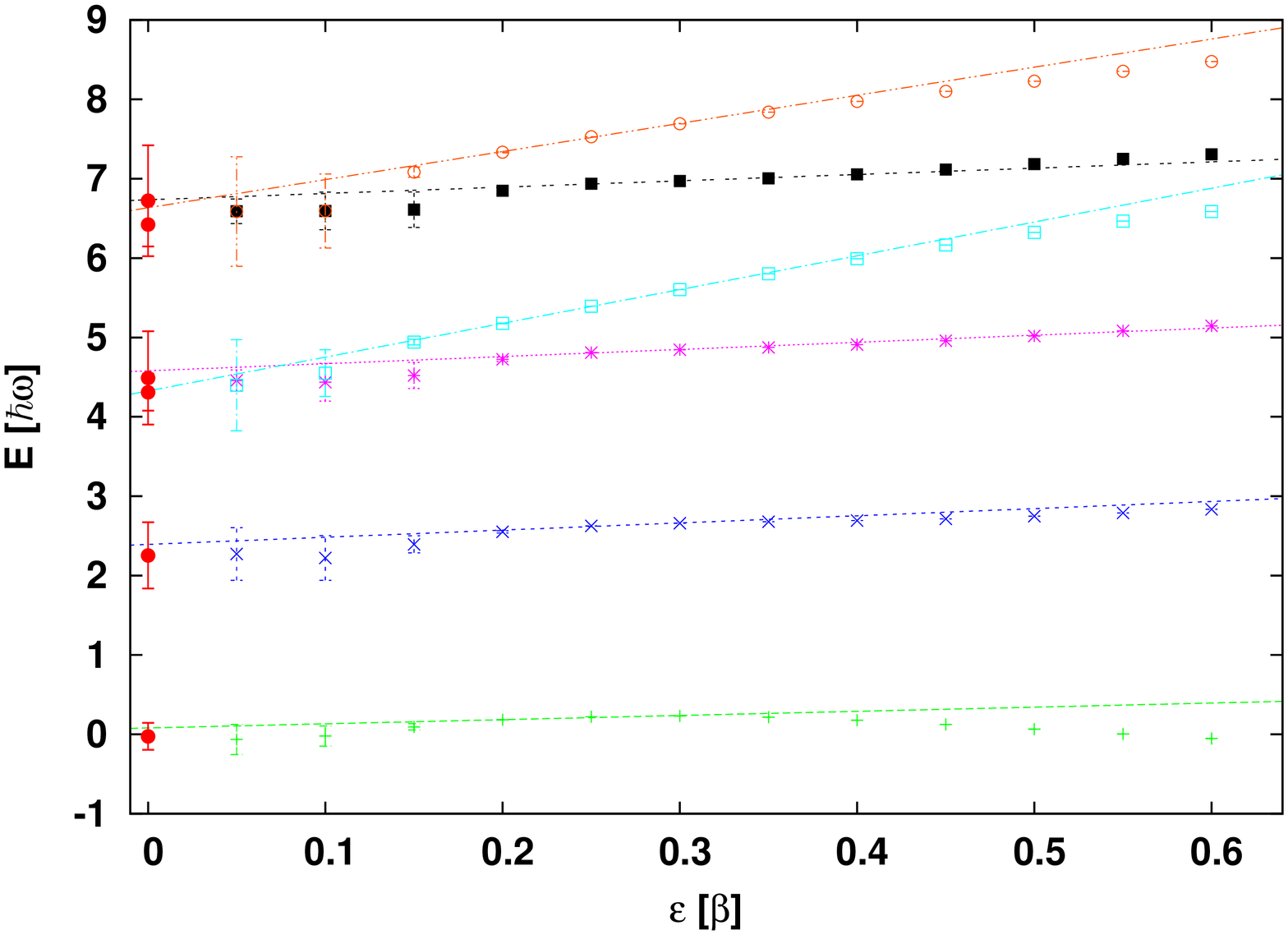}
  \caption{$L^{\pi}=0^+$ eigenvalues for $A=3$ as a function
    of $\epsilon$ with $E^{(2)}=-2$. The points at
    $\epsilon=0$ denote the extrapolated results for contact
    interactions.  The lines are linear extrapolations to
    $\epsilon=0$. Left panel: $E^{(3)}=-1$ Right panel:
    $E^{(3)}=-5$}
  \label{fig:3boson-2-5}
\end{figure}

Finally, we give in Fig.~\ref{fig:3boson-2-5} a non-unitary
example with $E^{(2)}=-2$. In the left panel the eigenvalues
for the renormalization energy $E^{(3)}=-1$ are shown as a
function of the width $\epsilon$ and in the right panel for
$E^{(3)}=-5$.  The results are linearly extrapolated like in
the unitary limit. At $\epsilon=0$ the results for contact
interactions extrapolated with a quadratic polynomial in
$1/\sqrt{N+3}$ are added. The uncertainties from the
extrapolation are estimated conservatively as the energy
shift from the last calculated eigenvalues to the
extrapolated one. Inside relative errors of $3.5\%$ referred
to the ground state with eigenenergy $-5$ respectively $-1$
both results coincide.
%%%%%%%%%%%%%%%%%%%%%%%%%%%%%%%%%%%%%%%%%%%%%%%%%%%%%%%%%%%%%%%%%%%%
%%%%%%%%%%%%%%%%%%%%%%%%%%%%%%%%%%%%%%%%%%%%%%%%%%%%%%%%%%%%%%%%%%%%
Our method clearly has larger errors for $E^{(3)}=-5$.

\subsection{Four identical bosons}
\label{sec:4identicalbosons}

\begin{figure}[ht]
  \centering
  \includegraphics[angle=270,
  width=0.49\linewidth]{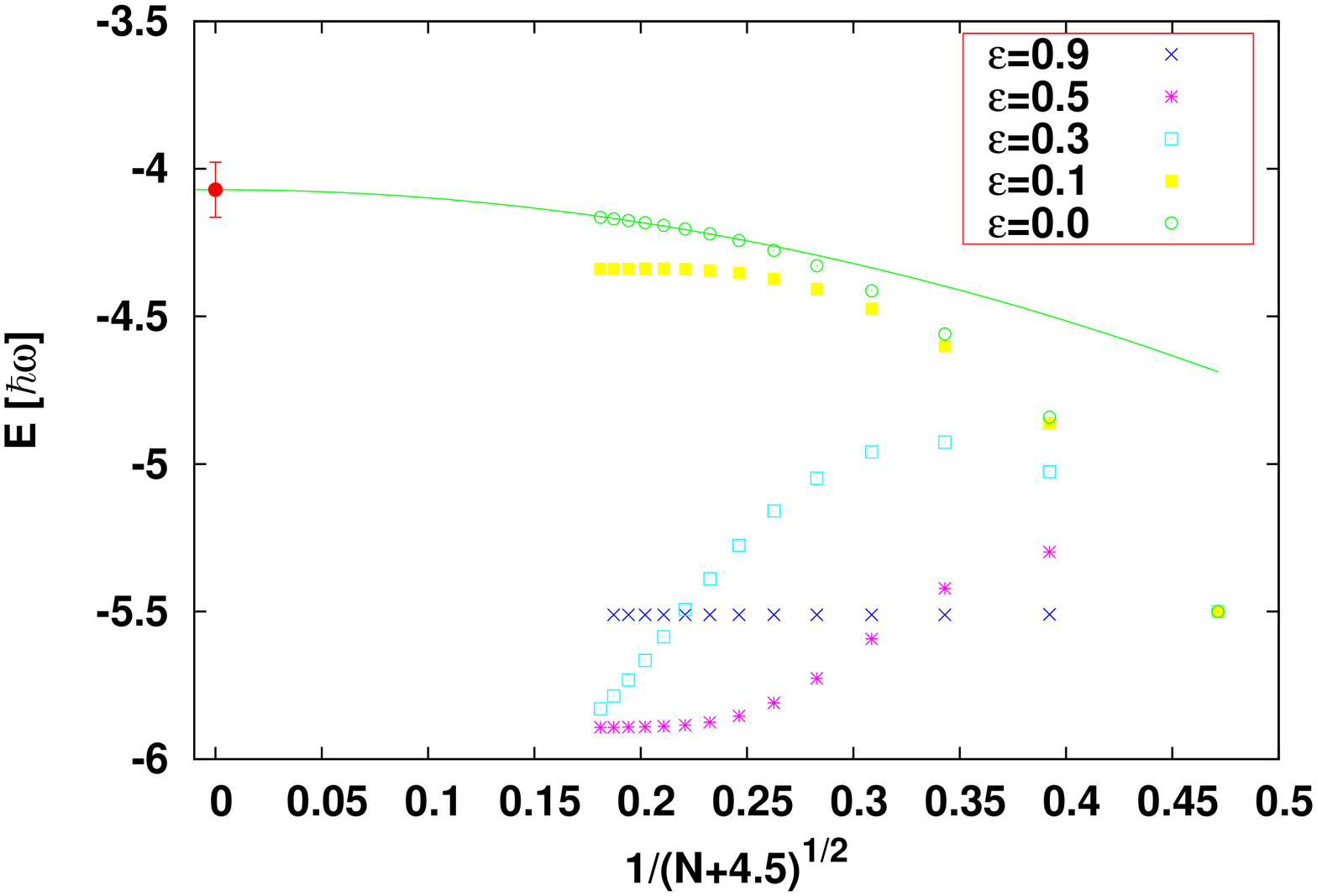}
  \includegraphics[angle=270,
  width=0.49\linewidth]{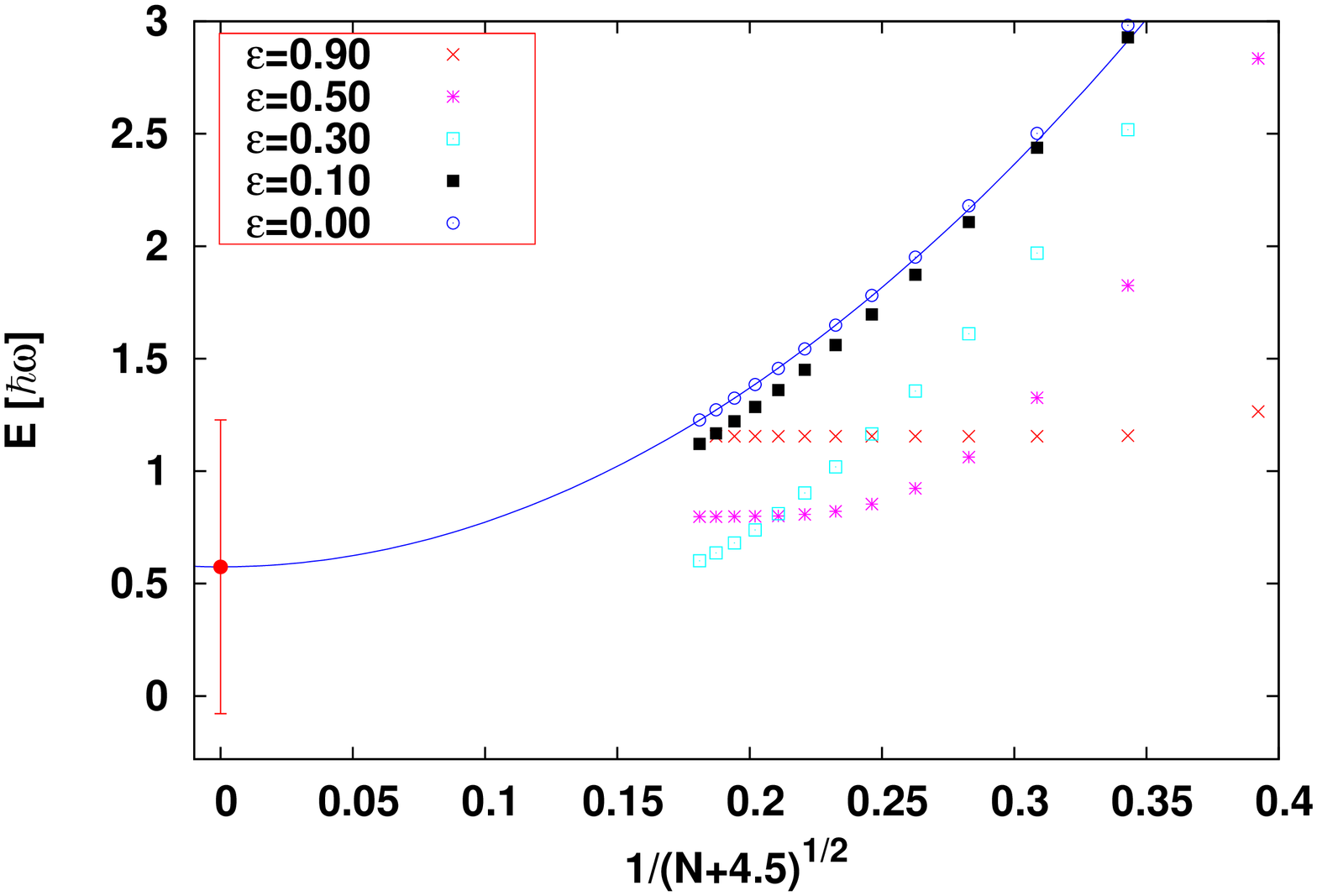}
  \caption{Four-boson $L^{\pi}=0^+$ states with
    $E^{(2)}=0.5$ and $E^{(3)}=-1$ for various smearing
    parameters $\epsilon$ as a function of
    $1/{\sqrt{N+4.5}}$. Left panel: Ground state. Right
    panel: First excited state. Solid lines are polynomial
    extrapolation.}
  \label{fig:4BL0+0.5-1run01}
\end{figure}

In Ref.~\cite{toelle2010}, we published results for a
4-boson system with contact interactions. Here we revisit
these calculations using smeared interactions. For instance,
calculations for the unitary limit $E^{(2)}=0.5$ with
$E^{(3)}=-1$ with contact interactions $\epsilon=0$ are
compared to calculations with $\epsilon\neq0$. We can reach
cutoff values up to $N=26$ which is significantly smaller
than in the three-body sector. In
Fig.~\ref{fig:4BL0+0.5-1run01}, the eigenenergies for the
ground and for the first excited states are shown for
various model-space sizes. The solid lines are
extrapolations with quadratic polynomials in
$1/{\sqrt{N+9/2}}$.
%%%%%%%%%%%%%%%%%%%%%%%%%%%%%%%%%%%%%%%%%%%%%%%%%%%%%%%%%%%%%%%%%%%%%%%%%
%%%%%%%%%%%%%%%%%%%%%%%%%%%%%%%%%%%%%%%%%%%%%%%%%%%%%%%%%%%%%%%%%%%%%%%%%%%%
As in the three-body sector the eigenvalues converge in $N$
earlier for larger smearing parameters $\epsilon$. However,
for smeared interactions the eigenvalues of the ground state
increase at first until they reach a maximum and begin to
decrease afterwards. It is conceivable that for contact
interactions and $\epsilon=0.1$ the model spaces are too
small in order to see the decrease of the eigenvalues. Thus,
our extrapolation given in the left panel of
Fig.~\ref{fig:4BL0+0.5-1run01} could have large systematic
errors. Since contact interactions can behave quite
differently from smeared interactions, however, a definite
answer requires calculations at larger cutoffs $N$ which are
beyond the scope of this work. If the renormalization
energies $E^{(2)}$ and $E^{(3)}$ are chosen equal, the
non-monotonous behavior in $\epsilon$ does not appear. It
becomes more severe as the mismatch between $E^{(2)}$ and
$E^{(3)}$ is increased.  The excited states do not show the
non-monotonous behavior in $\epsilon$ at all.  Their
eigenvalues decrease monotonously with $\epsilon$ and for
large $\epsilon$ they converge inside the model-space sizes
considered here.

\begin{figure}[ht]
  \centering
  \includegraphics[angle=270,
  width=0.49\linewidth]{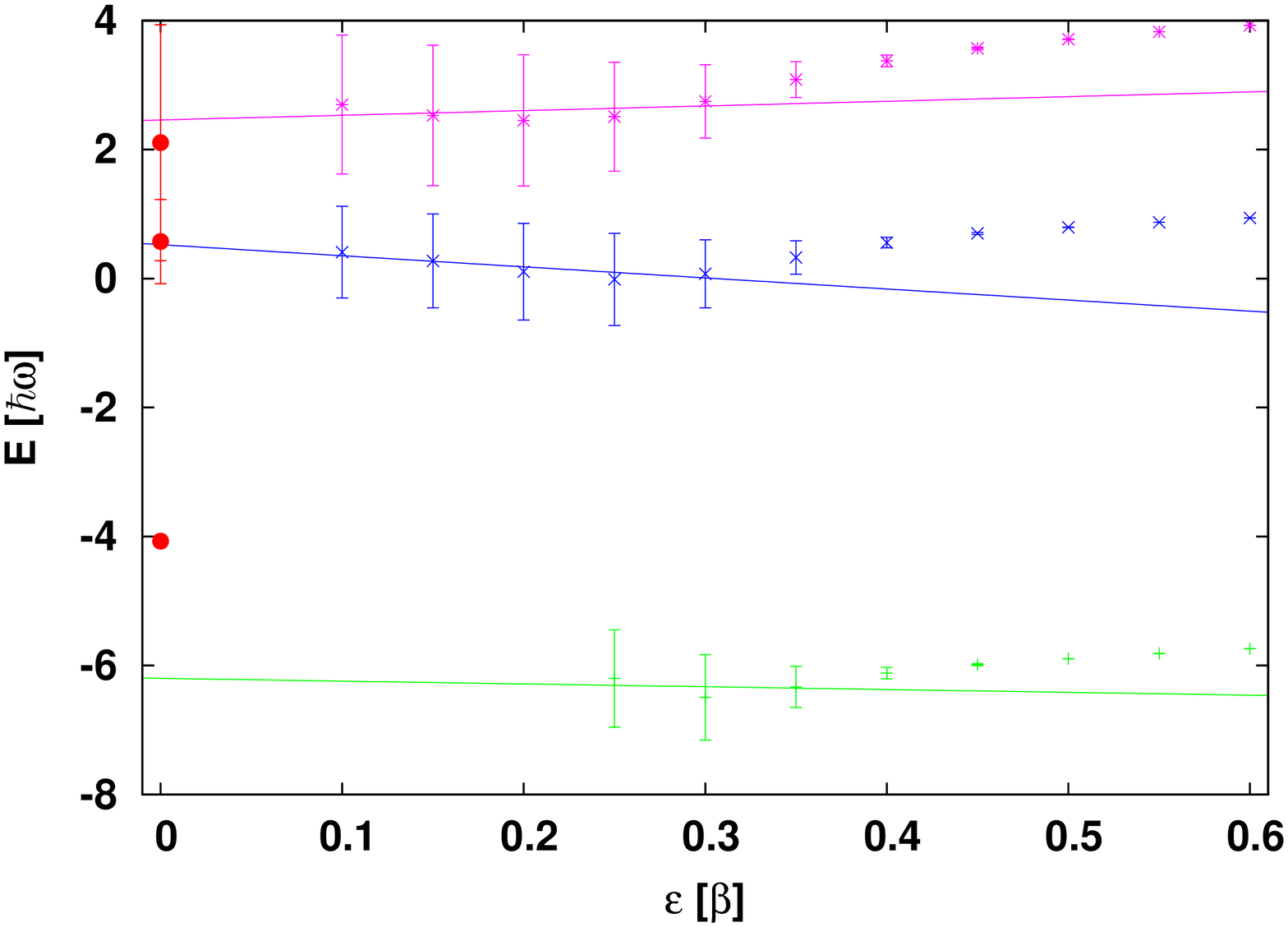}
  \includegraphics[angle=270,
  width=0.49\linewidth]{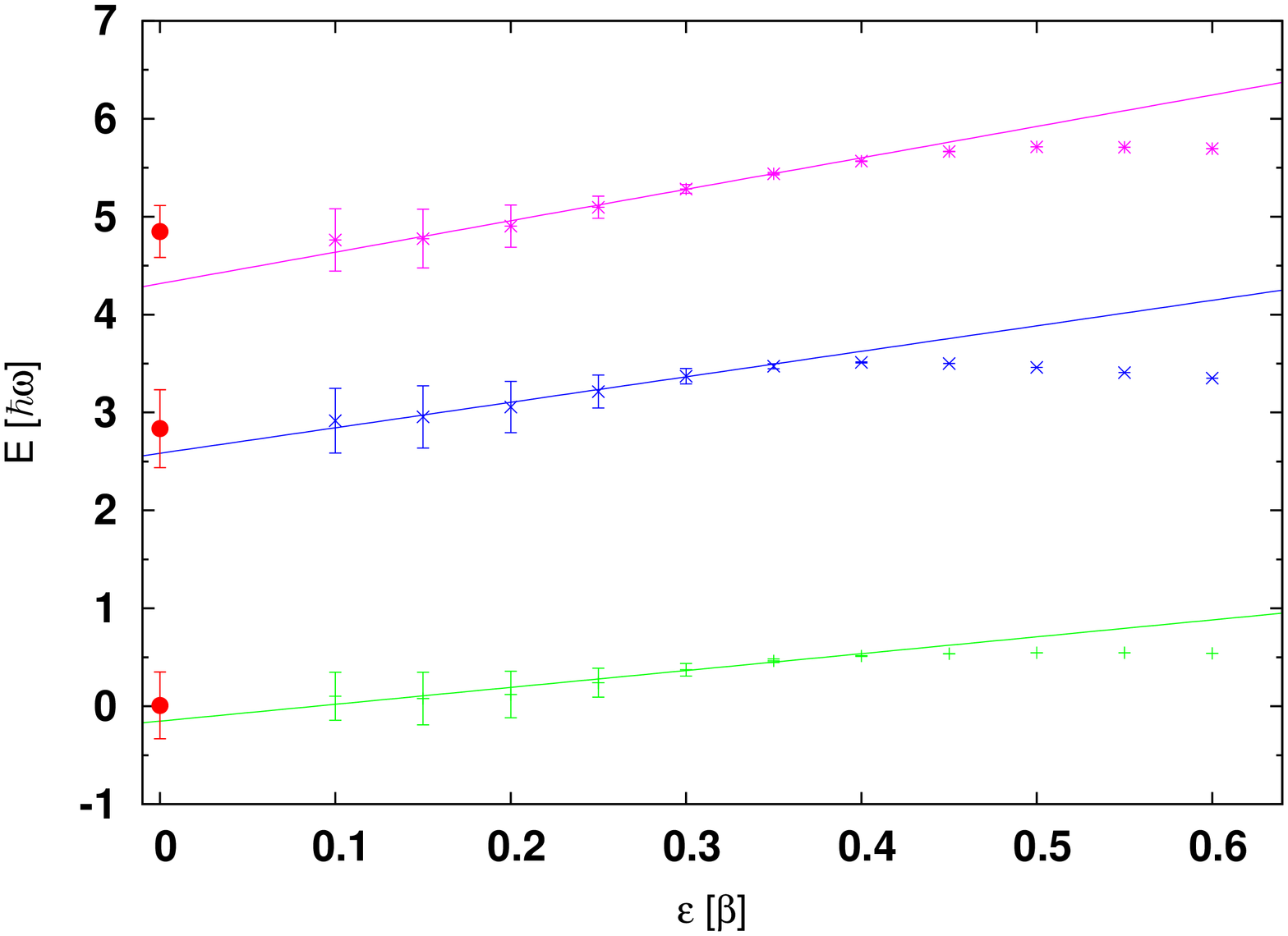}
  \caption{Lowest three energy states $L^{\pi}=0^+$ for
    $E^{(2)}=0.5$ as a function of $\epsilon$ for $A=4$.
    Left panel: $E^{(3)}=-1$. Right panel: $E^{(3)}=0.5$.
    The eigenvalues are linearly extrapolated. At
    $\epsilon=0$ extrapolated eigenvalues for contact
    interactions drawn in.}
  \label{fig:4BL02+0.5-1+0.5}
\end{figure}

In Fig.~\ref{fig:4BL02+0.5-1+0.5} (left panel) we show the
converged or extrapolated eigenvalues as a function of
$\epsilon$ for the three lowest states with renormalization
energies $E^{(2)}=0.5$ and $E^{(3)}=-1$. In comparison to
the three-body sector, we can diagonalize the Hamiltonian
only in small model spaces (up to $N=26$). Thus, just until
$\epsilon=0.35$ the eigenvalues have small error bars and we
extrapolated the states linearly to $\epsilon=0$. At
$\epsilon=0$ the eigenstates extrapolated via $N$ with a
quadratic polynomial in $1/\sqrt{N+9/2}$ and their estimated
uncertainties are shown. For $\epsilon<0.25$ the eigenvalues
for the ground states cannot be determined with acceptable
accuracy.  The extrapolated results differ significantly and
are at odds with each other for the ground state. This is
due to the non-monotonic behavior of the ground state energy
with $\epsilon$ discussed above.  For the excited states
this problem is not present and the results of the two
extrapolations are compatible.  In the right panel of
Fig.~\ref{fig:4BL02+0.5-1+0.5}, we show the results for
$E^{(3)}=0.5$. Since $E^{(2)}=E^{(3)}$, one expects less
uncertainties for a fixed $\epsilon$ than for the case
$E^{(2)} \neq E^{(3)}$ (see
section~\ref{sec:corr-due-scal}).
%%%%%%%%%%%%%%%%%%%%%%%%%%%%%%%%%%%%%%%%%%%%%%%%%%%%%%%%%%%%%%%%
%%%%%%%%%%%%%%%%%%%%%%%%%%%%%%%%%%%%%%%%%%%%%%%%%%%%%%%%%%%%%%
In particular, the ground state energies show no
non-monotonous dependence on $\epsilon$ and the
extrapolations in $\epsilon$ and $N$ are consistent.

 \subsection{5 and 6 identical bosons}
 We now apply our combined extrapolation technique to
 perform exploratory calculations of the spectra of 5 and 6
 identical bosons. In order to keep the uncertainties as
 small as possible, the renormalisations energies are chosen
 to be identical, i.e.  $E^{(2)}=E^{(3)}=0.5$. In
 particular, this removes the problem of the non-monotonous
 behavior of the energies as a function of $\epsilon$
 discussed above for the four-body system.  The cutoff
 parameter is $N=20$ for 5 bosons and $N=16$ for 6 bosons.

\begin{figure}[ht]
   \centering
   \includegraphics[angle=270,
   width=0.49\linewidth]{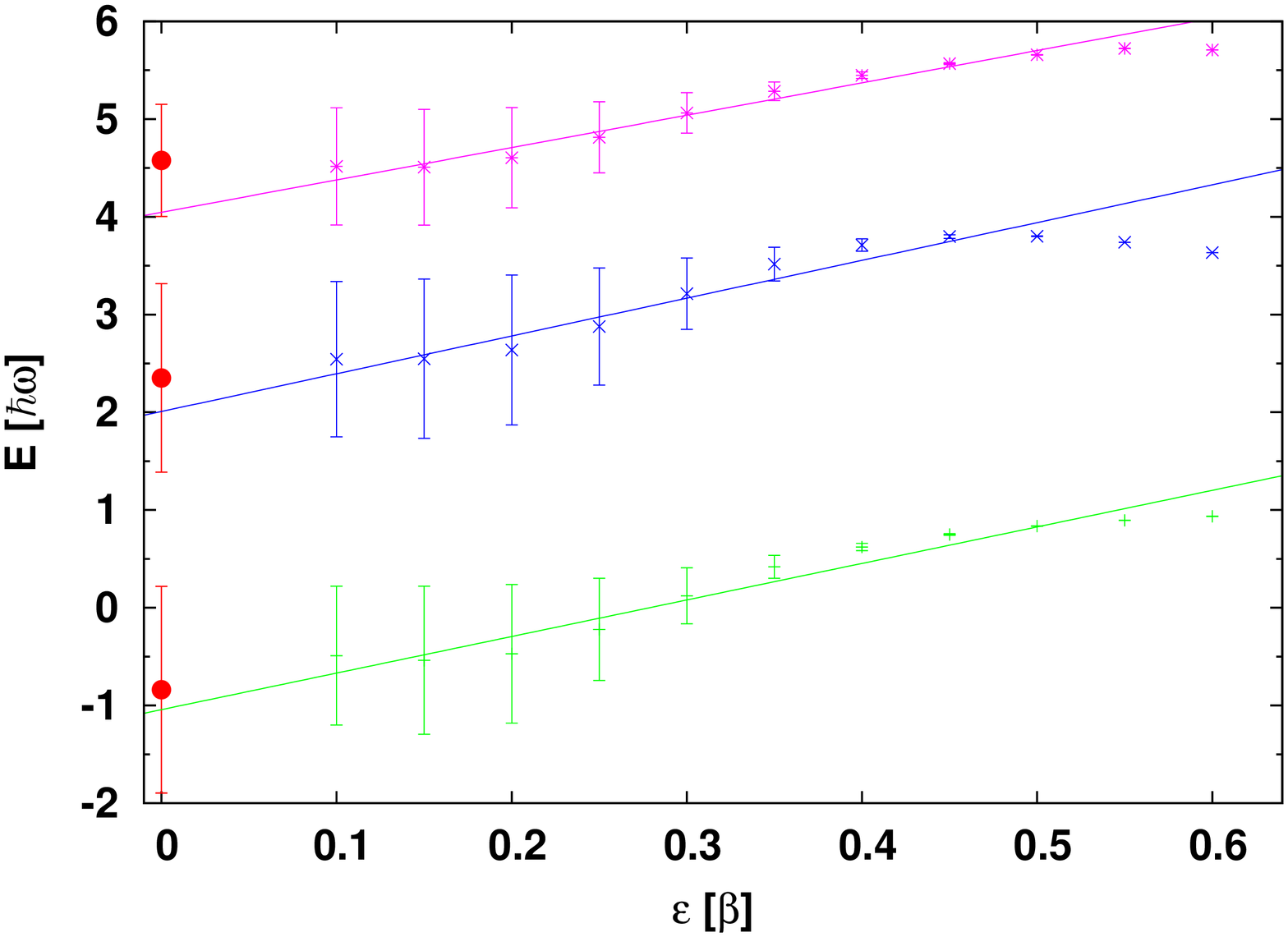}
   \includegraphics[angle=270,
   width=0.49\linewidth]{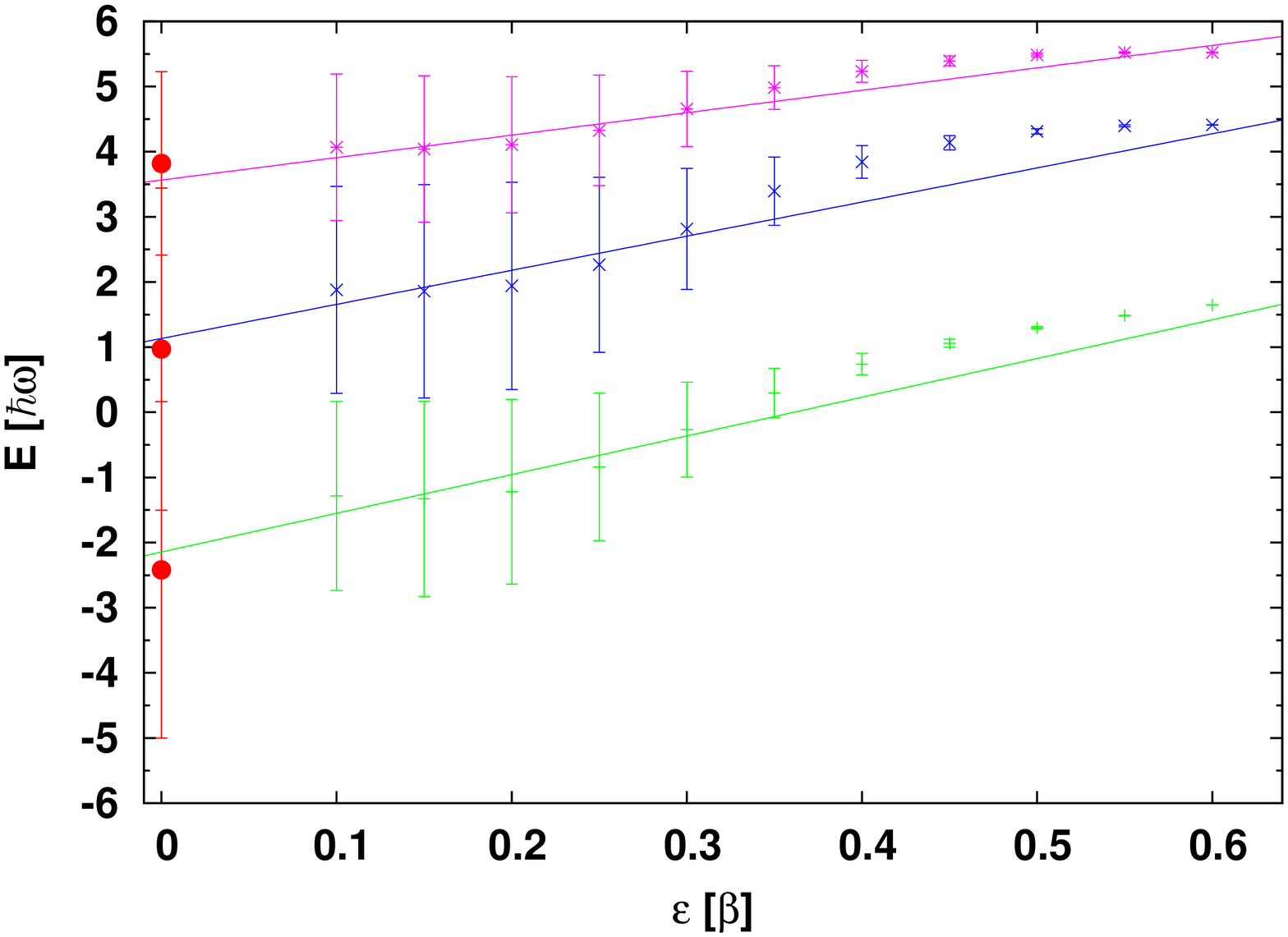}
   \caption{Lowest three energy states $L^{\pi}=0^+$ for
     $E^{(2)}=0.5$, $E^{(3)}=0.5$ as a function of
     $\epsilon$. Left panel: $A=5$ ($N=20$). Right panel:
     $A=6$ ($N=16$). The solid lines are linear
     extrapolations in $\epsilon$. At $\epsilon=0$ the
     eigenvalues extrapolated with a polynomial of second
     order in $1/\sqrt{N+(A-1)/2}$ for contact interactions
     drawn in.}
   \label{fig:56BL02+0.5-1}
 \end{figure}

 Our results for the three lowest energy $L^\pi = 0^+$
 states with different smearing parameters $\epsilon$ are
 shown in Fig.~\ref{fig:56BL02+0.5-1}.  Due to the small
 model spaces, the uncertainties are significantly larger
 than for three and four bosons.  The eigenenergies for
 contact interactions are extrapolated in $N$ with a
 polynomial of second order in $1/\sqrt{N+(A-1)/2}$.  They
 are shown at $\epsilon=0$ with conservative error
 bands. These approximated eigenvalues are consistent with
 extrapolations in $\epsilon$ for smeared contact
 interactions inside the error bands.
 %%%%%%%%%%%%%%%%%%%%%%%%%%%%%%%%%%%%%%%%%%%%%%%%%%%%%%%%%%%%%%%%%%%%%%%%
 %%%%%%%%%%%%%%%%%%%%%%%%%%%%%%%%%%%%%%%%%%%%%%%%%%%%%%%%%%%%%%%%%%%%%%%
 Thus we conclude that the combined extrapolation in $N$ and
 $\epsilon$ makes calculations in 5- and 6-boson systems
 with moderate computational resources possible.

 In Table~\ref{Tab:456res}, we have collected the extracted
 energies of the lowest three Efimov-like states for the
 unitary limit ($E^{(2)}=0.5$) and a three-body ground state
 energy fixed to $E^{(3)}=0.5$
 (cf.~Figs.~\ref{fig:3BGauss0.5-4},
 \ref{fig:4BL02+0.5-1+0.5}, and \ref{fig:56BL02+0.5-1}).
%%%%%%%%%%%%%%%%%%%%%%%%%%%%%%%%%%%%%%%%%%%%%%%%%%%%%%%%%%%%%%%%%%%%%%%
 \begin{table}[ht]
   \begin{tabular}{|c|c|c|c|c|}\hline
     $A=2$ & $A=3$ & $A=4$  &  $A=5$  & $A=6$ \\ \hline\hline
     0.5 & 0.5 & -0.1(2) & -0.9(2) & -2.3(3) \\
     2.5 & 2.9 &  2.7(3) &  2.2(3) & 1.1(2) \\
     4.5 & 5.1 &  4.6(5) &  4.3(5) & 3.7(3) \\\hline
   \end{tabular}
   \caption{Energies of the three lowest Efimov-like states in systems
     with $A=3,4,5,6$
     for the renormalization energies $E^{(2)}=0.5$ corresponding to
     the unitary limit and $E^{(3)}=0.5$.
     The column labeled $A=2$ contains the three lowest two-body states.
     All energies are in units of $\hbar\omega$.}
   \label{Tab:456res}
 \end{table}
%%%%%%%%%%%%%%%%%%%%%%%%%%%%%%%%%%%%%%%%%%%%%%%%%%%%%%%%%%%%%%%%%%%%%%%
 The numbers in the first two columns are the exact values
 for $A=2$ and $A=3$ rounded to two digits of accuracy. The
 values for the columns labeled $A=4$, $A=5$, and $A=6$ are
 extracted from our calculations. Here the number in
 parentheses gives the difference in the last digit between
 the extrapolation in $N$ and $\epsilon$ which can serve as
 an estimate of the numerical error.  The spectra are much
 more compressed than the corresponding free space
 results~\cite{Hammer:2006ct,Stecher:2008,Hanna:2006,Stecher:2011,Gattobigio:2011,Nicholson:2012}.
 However, a similar pattern can be observed: starting from
 $A=3$, every $A$-body state is accomponied by a
 corresponding $A+1$-body state. A second $A+1$-body state
 attached to the $A$-body state is not present in our
 calculation.  It would be interesting to study the spectra
 also away from the unitary limit for fixed $E^{(3)}$ to
 investigate their systematics. However, due to the
 additional errors from the mismatch in $E^{(2)}$ and
 $E^{(3)}$ this is numerically more challenging.

 \section{Conclusion and Outlook}
 \label{sec:sumconc}

 In this paper, we have extended our previous work on
 universal few-body physics in harmonic confinement
 \cite{toelle2010}.  We investigated $A$-body systems with
 up to six particles using genuine and smeared contact
 interactions at leading order in the large scattering
 length.

 The effective theory is defined within a finite model space
 with a cutoff $N$ on the basis functions. We perform a
 detailed study of the running of the effective two- and
 three-body coupling constants $g^{(2)}$ and $g^{(3)}$ with
 $N$ for contact interactions.  In the two-body sector, the
 results of Stetcu et al.~\cite{Stetcu:2010} for the running
 of $g^{(2)}$ are confirmed. For $g^{(3)}$, the limit cycle
 behavior expected from free space is observed and one
 complete cycle is traced numerically.

 Moreover, we have performed a detailed analysis of the
 leading effective theory errors using Lepage plots.  Our
 cutoff $N$ on the size of the model space implies both an
 infrared cutoff of order $\sqrt{N+(A-1)3/2}$ and an
 ultraviolet cutoff of order $1/\sqrt{N+(A-1)3/2}$ in the
 effective theory. The errors in the eigenenergies from
 these cutoffs for genuine contact interactions scale as
 $1/\sqrt{N+(A-1)3/2}$ in agreement with the expectations
 from the continuum case \cite{Bedaque:1998kg}.  For smeared
 contact interactions, we find an exponential dependence on
 $\sqrt{N}$ (cf.~Eq.~(\ref{eq:2})) in agreement with
 Ref.~\cite{Coon:2012}.

 An additional error is introduced if there is a mismatch in
 renormalization energies $E^{(2)}$ and $E^{(3)}$. We
 investigate these errors for contact interactions
 numerically and find that they are governed by the
 difference $(E^{(2)}-E^{(3)})$.  To improve convergence
 $E^{(2)}$ and $E^{(3)}$ should be chosen as close as
 possible.

 Smearing the contact interactions over distances of order
 $\epsilon$ improves the convergence with $N$ considerably.
 Obviously, this smearing changes the interaction.  In order
 to extract results for genuine contact interactions from
 calculations with finite $\epsilon$, extrapolations in
 $\epsilon$ are required. These extrapolations show a
 qualitatively different behavior for finite $N$ and
 infinite $N$ since the limits $N\to\infty$ and $\epsilon\to
 0$ do not commute.

 Based on our findings, the optimal strategy to extract the
 universal energies for contact interactions from numerical
 calculations in a finite model space when convergence can
 not be reached is to (i) use genuine contact interactions
 and extrapolate to $N=\infty$ and (ii) use converged
 energies for smeared contact interactions and extrapolate
 linearly to $\epsilon=0$. The difference between the two
 methods can be used to estimate the errors in the
 calculation.

 We use this strategy to calculate the spectra for $A$-boson
 systems with up to six bodies and study the case
 $E^{(2)}=E^{(3)}=0.5$ in detail.  We find that starting
 from $A=3$, every $A$-body state is accomponied by a
 corresponding $A+1$-body state. A second $A+1$-body state
 is not observed. The combined extrapolation in $N$ and
 $\epsilon$ makes the calculations possible with moderate
 computational resources and provides a first step towards
 calculating trapped many-body systems with resonant
 interactions. On the experimental side, the deterministic
 preparation of tunable few-fermion systems of up to ten
 particles in a microtrap was recently achieved by Serwane
 et al.~\cite{Serwane:2011}. Similar experimental studies of
 few-boson systems appear also promising.

 If there is a mismatch in $E^{(2)}$ and $E^{(3)}$, the
 ground state energies for four and more particles show a
 non-monotonous dependence on $N$ such that reliable results
 can only be obtained in large model spaces. For the excited
 states this problem is not present.  In the future, it
 would be interesting to get a better analytical
 understanding of these corrections and possibly develop an
 renormalization group improved matching procedure for this
 case.

 Finally, it would also be interesting to remove the
 trapping potential from the Hamiltonian (\ref{eq:effham})
 and apply our method to universal bound states in free
 space. For nuclear systems in the framework of the no-core
 shell model such a strategy was applied by Stetcu and
 collaborators~\cite{Stetcu:2006ey}.

\begin{acknowledgments}
  We thank S.~Coon, R.~Furnstahl, and U.~van Kolck for
  discussions and S.~Coon for pointing out previous work on
  the convergence properties of truncated basis expansions.
  This research was supported in part by the BMBF under
  contract No. 06BN7008.  H.W.H. acknowledges the INT
  program ``Light nuclei from first principles'' during
  which this work was finalized.
\end{acknowledgments}

\appendix
\section{Construction of the Hamiltonian Matrix}
\label{sec:hamiltonmatrixdetails}

In this appendix, we discuss the construction of the
Hamiltonian matrix for spinless bosonic wave functions,
i.e.\ symmetric functions under transposition of the
bodies. A symmetric basis for the model space of $A$ bodies
is constructed iteratively starting from the symmetric basis
for two bodies.

\subsubsection{Symmetric basis}\label{sec:symmetricbasis}

To construct unnormalised symmetric states we define
$P^{(2)}=(1+T_{12})$ with the transposition $T_{i k}$
between body $i$ and $k$. With the oscillator states
$\phi_{n l m}$ and the multi-index $u$ we thus have for 2
particles
\begin{align}
  P^{(2)}\phi_{n l m}(\vec{s}_1) &= \phi_{n l
    m}(\vec{s}_1)+\phi_{n l m}(-\vec{s}_1)=
  \begin{cases}
    0\quad&\textrm{for $l$ odd,}\\
    2\phi_{n l m}(\vec{s}_1)\quad&\textrm{for $l$ even,}
  \end{cases}=:
  \widetilde{\overline{\phi^{(2)}}}_{u}(\vec{s}_1)\,.
\end{align}
The orthonormal basis for 2 particles with multi-index
$\alpha$ thus is
\begin{align}
  \overline{\phi^{(2)}}_{\alpha}(\vec{s}_1)= \phi_{n l
    m}(\vec{s}_1)\quad\textrm{for $l$ odd.}
\end{align}
From these states coupled mixed-symmetric 3-body states are
constructed as
$\bigl[\overline{\phi^{(2)}}_{\alpha}\otimes\phi_{n
  l}\bigr]^L_M$. Since $[H,\vec{L}]=0$, the matrix elements
are independent of $M$ and this quantum number is
suppressed.
 
Accordingly, on the symmetric, unnormalised A-particle
states we have
\begin{displaymath}
  P^{(A)}=(1+T_{1A}+\cdots+T_{(A-1)A})P^{(A-1)}=
  (1+\sum_{i=1}^{A-2}T_{(A-1)i}T_{(A-1)A}T_{i(A-1)}+T_{(A-1)A})
  P^{(A-1)}\,. 
\end{displaymath}
Furthermore, we introduce the transformation matrix
\begin{align}
  C_{\textrm{sym}}^{(A-1)\rightarrow
    A}[z;\alpha',n'_2,l'_2,L']&=
  \bbraket{\bigl[\overline{\phi^{(A-1)}}_{\alpha'}
    \otimes\phi_{n'_2l'_2}\bigr]^{L'}}
  {\bigl[\widetilde{\overline{\phi^{(A)}}}_{z}\bigl]}\,,
\end{align}
with multi-indices $\alpha'$ and $z$ from the
mixed-symmetric $A$-body states
$\bigl[\overline{\phi^{(A-1)}}_{\alpha'}\otimes\phi_{n'_2l'_2}\bigr]^{L'}$
to the symmetric $A$-body states
$\widetilde{\overline{\phi^{(A)}}}_{z}$, given by
\begin{align}
  C_{\textrm{sym}}^{(A-1)\rightarrow
    A}&[z;\alpha',n'_2,l'_2,L']\nonumber\\
  &=\bbracket{\bigl[\overline{\phi^{(A-1)}}_{\alpha'}
    \otimes\phi_{n'_2l'_2}\bigr]^{L'}}
  {P^{(A)}}{\bigl[\overline{\phi^{(A-1)}}_{\alpha}
    \otimes\phi_{n_2l_2}\bigl]^L}\,,\\
  \label{eq:a_sym}
  &=\bbracket{\bigl[\overline{\phi^{(A-1)}}_{\alpha'}
    \otimes\phi_{n'_2l'_2}\bigr]^{L'}}
  {(1+(A-1)T_{(A-1)A})}{\bigl[\overline{\phi^{(A-1)}}_{\alpha}
    \otimes\phi_{n_2l_2}\bigl]^L}\,,
\end{align}
where $z=\{\alpha,n_2,l_2,L\}$.

Our goal is to construct an orthonormal basis of symmetric
$A$-body states $\overline{\phi^{(A)}}_{\omega}$. By
diagonalization of the symmetric norm matrix
\begin{align}
  \mathcal{N}[z';z]=\bbraket{\widetilde{\overline{\phi^{(A)}}}_{z}}
  {\widetilde{\overline{\phi^{(A)}}}_{z'}}\,,
\end{align}
the diagonal matrix $D$ of non-negative eigenvalues is
found:
\begin{align}
  D=O\cdot \mathcal{N}\cdot O^T\,.
\end{align}
Vanishing eigenvalues do not correspond to symmetric states
and respective eigenstates are eliminated. Thus, we find the
non-quadratic transformation matrix $B_{\textrm{sym}}$ from
mixed-symmetric $A$-body states to orthonormal symmetric
states:
\begin{align}
  B_{\textrm{sym}}^{(A-1)\rightarrow
    A}[\omega;\alpha',n_2',l_2'L']=\frac{1}{\sqrt{D[\omega;j]}}
  O[j;z]A_{\textrm{sym}}^{(A-1)\rightarrow
    A}[z;n_2',l_2'L']\,,
\end{align}
where a sum over $j$ and $z$ is implied. Accordingly, the
orthonormal basis of symmetric states is given by
\begin{align}
  \bket{\overline{\phi^{(A)}}_{\omega}}=
  B_{\textrm{sym}}^{(A-1)\rightarrow
    A}[\omega;\alpha,n_2,l_2,L]
  \bket{\bigl[\overline{\phi^{(A-1)}}_{\alpha}\otimes\phi_{n_2l_2}\bigr]^L}\,.
\end{align}

\subsubsection{Explicit calculation of $C_{\textrm{sym}}^{(A-1)\rightarrow A}$}
The explicit calculation of
$C_{\textrm{sym}}^{(A-1)\rightarrow A}$ is based on the
Talmi-Moshinsky-transformation and on a change in the
coupling scheme. With the operator
$B^{(A-2)\rightarrow(A-1)}_{\textrm{sym}}$
Eq. \eqref{eq:a_sym} yields
\begin{multline}
  \label{eq:4}
  C_{\textrm{sym}}^{(A-1)\rightarrow
    A}[\alpha,n_2,l_2,L;\alpha',n'_2,l'_2,L']=\\
  B_{\textrm{sym}}^{(A-2)\rightarrow(A-1)}[\alpha;\gamma,n_1,l_1,L_{\alpha}]
  B_{\textrm{sym}}^{(A-2)\rightarrow(A-1)}
  [\alpha';\gamma',n'_1,l'_1,L'_{\alpha'}]\\
  \bbracket{\bigl[\bigl[\overline{\phi^{(A-2)}}^{L'_{\gamma'}}_{\gamma'}
    \otimes\phi_{n'_1,l'_1}\bigr]^{L'_{\alpha'}}
    \otimes\phi_{n'_2l'_2}\bigr]^{L'}} {(1+(A-1)T_{(A-1)A})}
  {\bigl[\bigl[\overline{\phi^{(A-2)}}^{L_{\gamma}}_{\gamma}
    \otimes\phi_{n_1,l_1}\bigr]^{L_{\alpha}}
    \otimes\phi_{n_2l_2}\bigr]^{L}}\,.
\end{multline}
In order to calculate matrix elements of the operator
$T_{(A-1)A}$, a transformation to coordinates
$\vec{\lambda}$ and $\vec{\mu}$ is performed since
$T_{(A-1)A}$ is diagonal in $\vec{\lambda}$ and $\vec{\mu}$
(see Fig.~\ref{fig:n_bodya}):
%%%%%%%%%%%%%%%%%%%%%%%%%%%%%%%%%%%%%%%%%%%%%%%%%%%%%%%%%%%%%%%%%%%%%
\begin{figure}[htbp]
  \centering
  \includegraphics[width=0.8\linewidth]{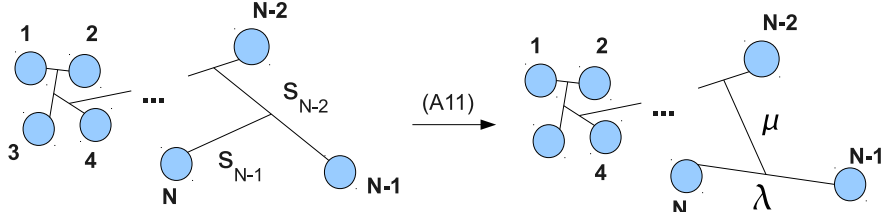}
  \caption{Talmi transformation from coordinates
    $\vec{s}_{\left(A-1\right)}$ and
    $\vec{s}_{\left(A-2\right)}$ to coordinates
    $\vec{\lambda}$ and $\vec{\mu}$}
  \label{fig:n_bodya}
\end{figure}
%%%%%%%%%%%%%%%%%%%%%%%%%%%%%%%%%%%%%%%%%%%%%%%%%%%%%%%%%%%%%%%%%%%%%
\begin{align}
  \begin{pmatrix}
    -\vec{\lambda}\\
    \vec{\mu}
  \end{pmatrix}
  =
  \begin{pmatrix}
    \sqrt{\frac{A-2}{2(A-1)}} & -\sqrt{\frac{A}{2(A-1)}}\\
    \sqrt{\frac{A}{2(A-1)}} & \sqrt{\frac{A-2}{2(A-1)}}
  \end{pmatrix}
  \cdot
  \begin{pmatrix}
    \vec{s}_{(A-2)}\\
    \vec{s}_{(A-1)}
  \end{pmatrix}\,.
\end{align}
For this purpose the coupling scheme is changed by means of
Wigner's 6j symbols \cite{varshalovich} and in the new
coupling scheme the Talmi-Moshinsky-transformation is
exploited.

With the Brody-Moshinsky-brackets
$\bbraket{n_{\lambda}l_{\lambda},n_{\mu}l_{\mu};L_{12}}
{n_1l_1,n_2l_2}_{\frac{A}{A-2}}$ we find for the wave
functions in terms of the coordinates $\vec{\lambda}$ and
${\vec{\mu}}$
\begin{multline}
  \Bigl[\bigl[\overline{\phi^{(A-2)}}^{L_{\gamma}}_{\gamma}
  \otimes\phi_{n_1,l_1}(\vec{s}_{\left(A-2\right)})\bigr]^{L_{\alpha}}
  \otimes\phi_{n_2l_2}(\vec{s}_{\left(A-1\right)})\Bigr]^{L}=\\
  \sum_{L_{12}}
  \left(-1\right)^{L_{\gamma}+l_1+l_2+L}\sqrt{\left(2L_{12}
      +1\right)\left(2L_{\alpha}+1\right)}
  \begin{Bmatrix}
    L_{\gamma} & l_1 & L_{\alpha} \\ l_2 & L& L_{12}
  \end{Bmatrix}\\
  \sum_{\stackrel{l_{\lambda},n_{\lambda}}{l_{\mu},n_{\mu}}}
  \Bigl(-1\Bigr)^{l_{\lambda}}
  \bbraket{n_{\lambda}l_{\lambda},n_{\mu}l_{\mu};L_{12}}
  {n_1l_1,n_2l_2}_{\frac{A}{A-2}}
  \Bigr[\overline{\phi^{\left(A-2\right)}}^{L_{\gamma}}_{\gamma}
  \otimes\bigl[\phi_{n_{\lambda}l_{\lambda}}(\vec{\lambda})
  \otimes\phi_{n_{\mu}l_{\mu}}(\vec{\mu})
  \bigr]^{L_{12}}\Bigr]^{L}\,.
\end{multline}

\subsubsection{Contributions of the 2- and 3- body contact
  interactions}

In addition to the matrix $C_{\textrm{sym}}$, also the
2-body interactions can be calculated in the coordinates
$\vec{\lambda}$ and $\vec{\mu}$ introduced in the last
section (see Fig.~\ref{fig:n_bodya}). Note that the
contributions of all 2-body interactions are equal since the
bodies are identical. The 2-body interaction between body
$(A-1)$ and $A$ depends only on the distance between these,
which is proportional to $\vec{\lambda}$ and therefore the
interaction is proportional to
$\delta^{(3)}(\vec{\lambda})$. With combinatorics the number
of interactions are determined: There are
% \begin{align}
%   \begin{pmatrix}
%     A\\
%     2
%   \end{pmatrix}
$ {A \choose 2} =\frac{A(A-1)}{2} $
% \end{align}
two-body interactions.

\begin{figure}[htbp]
  \centering
  \includegraphics[width=0.8\linewidth]{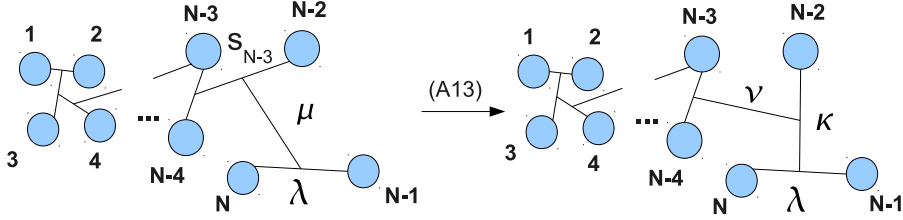}
  \caption{Talmi transformation from coordinates
    $\vec{s}_{\left(A-3\right)}$ and $\vec{\mu}$ to
    coordinates $\vec{\nu}$ and $\vec{\kappa}$}
  \label{fig:n_bodyb}
\end{figure}

Using the recoupling and the Talmi-Moshinsky-transformation
one more time the contributions of all
% \begin{align}
%   \begin{pmatrix}
%     A\\
%     3
%   \end{pmatrix}
$ {A\choose 3}= \frac{A(A-1)(A-2)}{6} $
% \end{align}
three-body interaction are determined in a similar fashion:
For states expressed in terms of the coordinates
$\vec{\nu}$, $\vec{\kappa}$ and $\vec{\lambda}$ the
interactions can be readily calculated (see
Fig.~\ref{fig:n_bodyb}). The corresponding transformation is
\begin{align}
  \begin{pmatrix}
    \vec{\nu}\\
    \vec{\kappa}
  \end{pmatrix}
  =
  \begin{pmatrix}
    \sqrt{\frac{A}{3(A-2)}} & -\sqrt{\frac{2(A-3)}{3(A-2)}}\\
    \sqrt{\frac{2(A-3)}{3(A-2)}} & \sqrt{\frac{A}{3(A-2)}}
  \end{pmatrix}
  \cdot
  \begin{pmatrix}
    \vec{s}_{(A-3)}\\
    -\vec{\mu}
  \end{pmatrix}\,.
\end{align}

\section{Smeared contact interaction}
\label{sec:smear-cont-inter}
In this section we give the derivation for the matrix
elements
\begin{align}
  \bbracket{n_1l_1}{V_G}{n_1'l_1'}&=
  \frac{1}{(1+\epsilon^2)^3} \phi_{n_10}(0)
  \Bigl(\frac{1-\epsilon^2}{1+\epsilon^2}\Bigr)^{n_1}
  \phi_{n_1'0}(0)
  \Bigl(\frac{1-\epsilon^2}{1+\epsilon^2}\Bigr)^{n'_1}
  \delta_{l_10}\delta_{l_1'0}\;.
\end{align}
For the interaction
\begin{equation}
  \bbracket{\vec{s}}{V_G}{\vec{s}'}= \frac{1}{(2\pi\epsilon^2)^3}
  e^{-\frac{\vec{s}^2}{2\epsilon^2}} e^{-\frac{\vec{s}'^2}{2\epsilon^2}}\,,
\end{equation}
we have
\begin{align}
  \bbracket{n_1l_1}{V_G}{n_1'l_1'}&=
  \frac{1}{(2\pi\epsilon^2)^3} \int\d^3s~\phi_{n l}(\vec{s})
  e^{-\frac{\vec{s}^2}{2\epsilon^2}} \int\d^3s'~\phi_{n
    l}(\vec{s}') e^{-\frac{\vec{s}'^2}{2\epsilon^2}}\;.
\end{align}
Moreover,
\begin{align}
  \int\d^3s~\phi_{n l}(\vec{s})
  e^{-\frac{\vec{s}^2}{2\epsilon^2}}&=
  \sqrt{4\pi}\int_{0}^{\infty}\d
  r~r^2e^{-\frac{r^2}{2\epsilon^2}} R_{n0}(r)= N_{n0}
  \int_{0}^{\infty}\d r~r^2
  e^{-\frac{1}{2}(1+\frac{1}{\epsilon^2})r^2}
  L_n^{(\frac{1}{2})}(r^2)\;,
\end{align}
with the associated Laguerre polynomials
$L_n^{(\frac{1}{2})}$ and the normalization $N_{n0}=
\sqrt{\frac{1}{\sqrt{4\pi}} \frac{2^{n+3}n!}{(2n+1)!!}}$. By
the substitution of $r^2=x$ and with
$\rho^2=\frac{1}{2}(1+\frac{1}{\epsilon^2})$, we get with
the expansion of the Laguerre polynomials in monomials
\begin{align}
  \int\d^3s~\phi_{n l}(\vec{s})
  e^{-\frac{\vec{s}^2}{2\epsilon^2}}&
  =\frac{\sqrt{4\pi}N_{n0}}{2}\sum_{i=0}^{n}
  \frac{(-1)^i}{i!}
  \frac{\Gamma(n+\frac{3}{2})}{\Gamma(i+\frac{3}{2})\Gamma(n-i+1)}
  \int_{0}^{\infty}\d x~x^{\frac{1}{2}+i}e^{-\rho^2x}\\
  &= \frac{\sqrt{4\pi}N_{n0}}{2}\sum_{i=0}^{n}
  \frac{(-1)^i}{i!}
  \frac{\Gamma(n+\frac{3}{2})}{\Gamma(n-i+1)} \frac{1}{\rho^{3+2i}}\\
  &=\frac{\sqrt{4\pi}N_{n0}\Gamma(n+\frac{3}{2})}{2\rho^3}\frac{1}{n!}
  \sum_{i=0}^{n} \binom{n}{i}(-\rho^{-2})^i\\
  &=\frac{\sqrt{4\pi}N_{n0}(2n+1)!!}
  {2^{n+2}n!\rho^{2n+3}}(\rho^2-1)^n\sqrt{\pi}\,,
\end{align}
and replacing $\rho$ and $N_{n0}$, we find
\begin{align}
  \bbracket{n_1l_1}{V_G}{n_1'l_1'}&=\frac{1}{\bigl(1+\epsilon^2\bigr)^{3}\pi^{\frac{3}{2}}}
  \sqrt{\frac{(2n_1+1)!!}{n_1!2^{n_1}}}
  \sqrt{\frac{(2n_1'+1)!!}{n_1'!2^{n_1'}}}
  \left(\frac{1-\epsilon^2}{1+\epsilon^2}\right)^{n_1}
  \left(\frac{1-\epsilon^2}{1+\epsilon^2}\right)^{n_1'}\\
  &=\frac{1}{(1+\epsilon^2)^3} \phi_{n_10}(0)
  \Bigl(\frac{1-\epsilon^2}{1+\epsilon^2}\Bigr)^{n_1}
  \phi_{n_1'0}(0)
  \Bigl(\frac{1-\epsilon^2}{1+\epsilon^2}\Bigr)^{n'_1}
  \delta_{l_10}\delta_{l_1'0}\;.
\end{align}

\section{Effective range expansion for smeared contact
  interactions}
\label{sec:ERE_SGI}
For separable potentials
\begin{align}
  \bbracket{\vec{s}}{V}{\vec{s}'}=
  g^{(2)}\omega(\vec{s})\omega(\vec{s}')\,,
\end{align}
we get for the T-matrix
\begin{align}
  \bbracket{\vec{p}}{T(z)}{\vec{p}'}&=
  \left(\frac{1}{g^{(2)}}-B(z)\right)^{-1}
  v(\vec{p})v(\vec{p}') & &\textrm{with} &
  B(z)&:=\int\frac{\d^3q}{(2\pi)^3}
  \frac{\left|v(\vec{q})\right|^2}{z-\frac{q^2}{2}}
\end{align}
from the Lippmann-Schwinger equation
\begin{align}
  T(z)=V+V(z-H_o)^{-1}T(z)\;.
\end{align}
Here, $v(\vec{p})$ denotes the Fourier transform of
$\omega(\vec{s})$
\begin{align}
  v(\vec{p})=\int\d^3s~e^{i\vec{p}\vec{s}}\omega(\vec{s})\;.
\end{align}

For separable central potentials there is only s-wave
scattering and we find for the scattering amplitude $f_0(p)$
with the effective range expansion
\begin{align}
  f_0(p)&=\frac{1}{-\frac{1}{a}+\frac{1}{2}r p^2+
    \mathcal{O}(p^4)-i p}\\
  &=-\frac{1}{2\pi}\lim_{\Delta\rightarrow0}
  \bbracket{\vec{p}}{T(p^2/2+i\Delta)}{\vec{p}'}\\
  &=\left[-2\pi|v(p)|^{-2} \bigl(1/g^{(2)}-
    B(p^2/2+i\Delta)\bigr)\right]^{-1}\,.
\end{align}

For the smeared contact interaction we have
\begin{align}
  \omega(\vec{s})&=\frac{1}{(2\pi\epsilon^2)^{\frac{3}{2}}}
  e^{-\frac{\vec{s}^2}{2\epsilon^2}}\;, &
  v(\vec{p})&=e^{-\frac{\vec{p}^2\epsilon^2}{2}}\;,
\end{align}
\begin{displaymath}
  B(p^2/2+i\Delta)=\frac{1}{\pi^2}
  \int_{0}^{\infty}\d q~\frac{q^2e^{-\epsilon^2q^2}}{p^2-q^2+i\Delta}
  =\frac{1}{2\pi^2}\int_{0}^{\infty}\d
  k~\frac{\sqrt{k}e^{-\epsilon^2k}}{p^2-k+i\Delta}\;, 
\end{displaymath}
and the imaginary part of the denominator in $f_0(p)$ can be
calculated with
\begin{align}
  \lim_{\Delta\rightarrow 0^+} \int_a^b\d x~\frac{f(x)}{x\pm
    i \Delta} = \mp i \pi f(0) + \mathcal{P}\int_a^b \d
  x~\frac{f(x)}{x}\;.
\end{align}
It follows for the imaginary part
\begin{align}
  \Im{\frac{1}{f_0}}= -\pi 2\pi
  e^{+\epsilon^2p^2}\frac{1}{2\pi^2}p e^{-\epsilon^2p^2}=-p
\end{align}
and for the real part
\begin{align}
  \Re{\frac{1}{f_0}}=\left[-\frac{2\pi}{g^{(2)}}
    -\frac{1}{\pi}\underbrace{\left[\mathcal{P}
        \int_{-\infty}^{+\infty}\d
        q~\frac{q^2e^{-\epsilon^2q^2}}{q^2-p^2}\right]}_{A(p^2)}
  \right] e^{\epsilon^2p^2}\:.
\end{align}
Moreover, it can be shown, that
\begin{align}
  A(p^2)=\sqrt{\pi}\frac{1-2\epsilon p \textrm{F}(\epsilon
    p)}{\epsilon}\;,
\end{align}
where the Dawson function is defined by
\begin{align}
  \label{eq:DawsonF}
  \textrm{F}(x)=e^{-x^2}\int_{0}^{x}\d y~e^{y^2}
  =x-\frac{2}{3}x^3+\mathcal{O}(x^5)\;.
\end{align}
 
The function $A(p^2)$ as well $e^{\epsilon^2p^2}$ is
expanded in $p^2$. We will be interested in the scattering
length $a$ and the effective range $r$, thus the expansion
is terminated at $\mathcal{O}(p^4)$:
\begin{align}
  -\frac{1}{a}+\frac{1}{2}r p^2+\mathcal{O}(p^4)&=
  \left[-\frac{2\pi}{g^{(2)}} -
    \frac{1}{\pi}\frac{\sqrt{\pi}}{\epsilon}
    \biggl(1-2\epsilon p\Bigl(\epsilon p-
    \frac{2}{3}\epsilon^3p^3\Bigr)\biggr)\right]
  \biggl(1+\epsilon^2p^2\biggr)+\mathcal{O}(p^4)\;,\\
  \label{eq:ere-smeared-int}
  &=-\bigl(\frac{2\pi}{g^{(2)}}+
  \frac{1}{\sqrt{\pi}\epsilon}\bigr)+
  \frac{1}{2}\left(\frac{2\epsilon}{\sqrt{\pi}}
    -\frac{4\pi\epsilon^2}{g^{(2)}} \right)p^2+
  \mathcal{O}(p^4)\;.
\end{align}
Note, that the momentum $\vec{p}$ is the Jacobi
momentum. The relation between the oscillator length for
canonical relative coordinates and Jacobi coordinates is
just a factor of $\sqrt{2}$ in the two-body sector and the
oscillator lengths differ with the same factor. Therefore
all lengths must be rescaled with $\sqrt{2}$ in order to get
the canonical effective range expansion.

\end{document}